\documentclass{aastex63}

\usepackage{newtxtext,newtxmath}
\usepackage[T1]{fontenc}
\usepackage{ae,aecompl}
\usepackage{graphicx}
\usepackage{amsmath}
\usepackage{gensymb}
\usepackage{amssymb}
\usepackage{color}
\usepackage{subfiles}
\usepackage[british]{babel}
\usepackage[utf8]{inputenc}
\definecolor{comment}{RGB}{200,0,0}

\received{\today}
\accepted{\today}
\submitjournal{ApJ}

\shorttitle{Galaxy Zoo Builder: Method paper}
\graphicspath{{./}{figures/}}

\begin{document}

\title{Galaxy Zoo Builder: Four Component Photometric decomposition of Spiral Galaxies Guided by Citizen Science}

\correspondingauthor{Timothy K. Lingard}
\email{tklingard@gmail.com}

\author[0000-0002-0529-7427]{Timothy K. Lingard}
\affiliation{Institute of Cosmology and Gravitation, University of Portsmouth \\
Dennis Sciama Building, Burnaby Road, Portsmouth, PO1 3FX, UK}

\author[0000-0003-0846-9578]{Karen L. Masters}
\affiliation{Haverford College, 370 Lancaster Ave., Haverford, PA 19041, USA}

\author[0000-0001-9233-2341]{Coleman Krawczyk}
\affiliation{Institute of Cosmology and Gravitation, University of Portsmouth \\
Dennis Sciama Building, Burnaby Road, Portsmouth, PO1 3FX, UK}

\author[0000-0001-5578-359X]{Chris Lintott}
\affiliation{Oxford Astrophysics, Denys Wilkinson Building, Keble Road, Oxford, OX1 3RH, UK}

\author[0000-0001-8010-8879]{Sandor Kruk}
\affiliation{European Space Agency, ESTEC, Keplerlaan 1, NL-2201 AZ, Noordwijk, The Netherlands}

\author[0000-0001-5882-3323]{Brooke Simmons}
\affiliation{Physics Department, Lancaster University, Lancaster, LA1 4YB, UK}

\author[0000-0003-2896-2503]{Robert Simpson}
\affiliation{Google UK, Six, Pancras Square, London N1C 4AG}

\author[0000-0001-7821-7195]{Steven Bamford}
\affiliation{Centre for Astronomy \& Particle Theory, School of Physics \& Astronomy \\
University of Nottingham, Nottingham, NG7 2RD, UK}

\author[0000-0003-0939-6518]{Robert C. Nichol}
\affiliation{Institute of Cosmology and Gravitation, University of Portsmouth \\
Dennis Sciama Building, Burnaby Road, Portsmouth, PO1 3FX, UK}

\author{Elisabeth Baeten}
\affiliation{Independent Zooniverse Volunteer}



\begin{abstract}
  Multi-component modelling of galaxies is a valuable tool in the effort to quantitatively understand galaxy evolution, yet the use of the technique is plagued by issues of convergence, model selection and parameter degeneracies. These issues limit its application over large samples to the simplest models, with complex models being applied only to very small samples. We attempt to resolve this dilemma of ``quantity or quality'' by developing a novel framework, built inside the Zooniverse citizen science platform, to enable the crowdsourcing of model creation for \replaced{SDSS}{Sloan Digitial Sky Survey} galaxies. We have applied the method, including a final algorithmic optimisation step, on a test sample of 198 galaxies, and examine the robustness of this new method. We also compare it to automated fitting pipelines, demonstrating that it is possible to consistently recover accurate models that either show good agreement with, or improve on, prior work. We conclude that citizen science is a promising technique for modelling images of complex galaxies, and release our catalogue of models.
\end{abstract}

\section{Introduction}
\label{sec:introduction}

Disc galaxies are complex objects, containing many different components, including a disc, disc phenomena (i.e. spiral arms, bars and rings) and central structures (bulges, bars). Decomposing disc galaxies into their component structures has become an important tool for extragalactic astronomers seeking to understand the formation and evolution of the galaxy population, ranging from analysing bulge and bar structure (\citealt{1985ApJ...288..438E}, \citealt{1996A&AS..118..557D}, \citealt{2011MNRAS.415.3308G}, \citealt{Mendez-Abreu2016:1610.05324v1}, \citealt{Gao2017:1709.00746v1}, \citealt{2018MNRAS.473.4731K}) to the secular evolution of disc galaxies (\citealt{1998ApJ...500...75L}, \citealt{2005ApJ...635..959B}) and general galaxy assembly and evolution (\citealt{Simard2002:astro-ph/0205025v2}, \citealt{megamorph-paper}, \citealt{2012MNRAS.421.2277L}, \citealt{2019arXiv191002664R}).

These fully quantitative methods allow researchers to obtain structural parameters of galaxy sub-components, which has been used in a variety of astrophysical and cosmological research. For example, the stellar mass found in discs and bulges places strong constraints on the galaxy merger tree from $\Lambda$CDM $N$-body simulations (\citealt{2009MNRAS.396.1972P}, \citealt{Hopkins2010:1004.2708v3}, \citealt{2018MNRAS.475.5133R}); the strength of a galaxy's classical bulge is thought to be tied to the strength of merger events in its past (\citealt{2005ApJ...622L...9S}, \citealt{Kormendy2010:1009.3015v1}); different spiral arm formation theories vary in their predictions of spiral morphology (\citealt{Dobbs2014:1407.5062v1}, \citealt{Pour-Imani2016:1608.00969v1}, \citealt{2017MNRAS.472.2263H}).

The usefulness of obtaining parametric models of a galaxy has motivated the creation of many image modelling and fitting suites, including \textsc{Gim2d} \citep{Simard2002:astro-ph/0205025v2}, \textsc{Galfit} \citep{galfit-paper}, \textsc{MegaMorph} \citep{megamorph-paper} and \textsc{Profit} \citep{profit-paper} to name a few. Using these tools, researchers have built large catalogues of model fits to galaxies. One of the largest photometric model catalogues is that of \citet{2011ApJS..196...11S}, who performed automated 2D, two-component (bulge + disc) decomposition of 1,123,718 galaxies from the Legacy imaging of the Sloan Digital Sky Survey (hereafter SDSS) Data Release 7 \citep{SDSSDR7}.

Many other large catalogues of photometric fits exist (i.e. \citealt{2012MNRAS.421.2277L}, \citealt{2012MNRAS.421.1007K}, \citealt{2012ApJS..203...24V}), but despite the usefulness of photometric fitting, and the presence of analytic profiles and methods for modelling more complex galaxy sub-components, relatively few studies have attempted to perform large-scale (1000s of galaxies) parametric decomposition of galaxies using more complicated models than that of \citet{2011ApJS..196...11S} (two axisymmetric S\'ersic components representing a bulge and disc). Not properly taking into account these ``secondary'' morphological features (such as a bar, ring and spiral arms) can impact detailed measurements of a galaxy's bulge \citep{Gao2017:1709.00746v1}. Proper decomposition of secondary morphological features allows investigation into mechanisms behind the secular evolution of galaxies (\citealt{2015MNRAS.453.3729H}, \citealt{2018MNRAS.473.4731K}, \citealt{2018ApJ...862..100G}) and exploration of environmental effects on morphology, such as offset bars \citep{2017MNRAS.469.3363K}.

A prominent issue when performing these detailed decompositions is the tendency for fitting functions to converge on unphysical results when not properly guided or constrained, for instance in a two-component model containing a S\'ersic bulge and an exponential disc, the bulge may grow to encompass the galaxy's disc, as its extra parameter allows for a closer fit (as observed by \citealt{2018MNRAS.473.4731K}). It is also the case that often, without near-optimal starting points, detailed model fits will fail to converge at all \citep{2016MNRAS.462.1470L}.

Compounding this, uncertainties reported by many software fitting packages (i.e. \textsc{Galfit} and \textsc{MegaMorph} from the above list) are often lower estimates on the real uncertainty, due to secondary sources not being modelled, flat-fielding errors and incorrect models (including the possibly incorrect assumption of Poisson noise) \citep{2010AJ....139.2097P}. Other packages such as \textsc{Gim2d} and \textsc{Profit} attempt to fully model posterior distributions and so produce more representative uncertainties, however this comes with a larger computational cost and configuration complexity. Formal uncertainties are measures of the likelihood space and therefore underestimates of the true error as an analytic model will rarely capture the nuanced light profile of a galaxy.

Another problem that needs to be addressed is whether a component should be present in the model at all. An automated fit will generally attempt to add as many components as possible to produce the closest-matching model. Many studies therefore need to select the most appropriate model by visual inspection of the resulting residuals or recovered parameters. For example, both \citet{Vika2014:1408.4070v1} and \citet{2018MNRAS.473.4731K} inspected the resulting model and residual images for all of their parametric fits (163 and 5,282 respectively) to ensure physical results with the correct components present. The end result of most of these problems is that researchers will have to invest time to individually check many of their fits to ensure they have converged on a physical model. In the era of large sky surveys such as the SDSS, the time required to do this becomes unsustainable and introduces concerns over human error if done by only a single, or small number of individuals.

A demonstrably successful solution to the similar problem of galaxy classification in the era of large surveys, was to find a new source of person-power: \cite{Lintott2008:0804.4483v1} invited large numbers of people to classify SDSS-images of galaxies over the internet in the Galaxy Zoo project. The resulting classifications (a mean of 38 per galaxy) were then weighted and averaged to create a morphological catalogue of 893,212 galaxies. This hugely successful project, including its subsequent iterations and expansions (i.e. \citealt{Willett2013:1308.3496v2}, \citealt{Hart2016:1607.01019v1}, \citealt{2017MNRAS.464.4176W}, \citealt{2017MNRAS.464.4420S}), has produced a large catalogue of detailed morphological classifications that are in good agreement with other studies, and have been used in a wide variety of studies of the local galaxy population (see \citealt{2019arXiv191008177M} for a recent review).

In this paper we explore an analogous solution to that \citet{Lintott2008:0804.4483v1} brought to galaxy classification for the issues faced by galaxy parametric modelling, inside the ecosystem that Galaxy Zoo set in motion (namely {\it The Zooniverse}\footnote{\url{https://www.zooniverse.org}}). We leverage citizen scientists to pick model components and perform model optimization in an online, web-browser environment\footnote{\url{https://www.zooniverse.org/projects/tingard/galaxy-builder}}. We describe our method in Section \ref{sec:method}, including details of the images and ancillary data from SDSS as well as the strategy used to obtain scientifically useful models from volunteer classifications. We provide consistency checks within our infrastructure and to other methods in Section \ref{sec:results} and discuss the efficacy and potential impact of our new method relative to existing methodologies in Section \ref{sec:conclusions}.

Where necessary, we make use of $H_0 = 70\ \text{km}\ \text{s}^{-1}\ \text{Mpc}^{-1}$.

\section{Method}
\label{sec:method}
\begin{figure*}
  \plotone{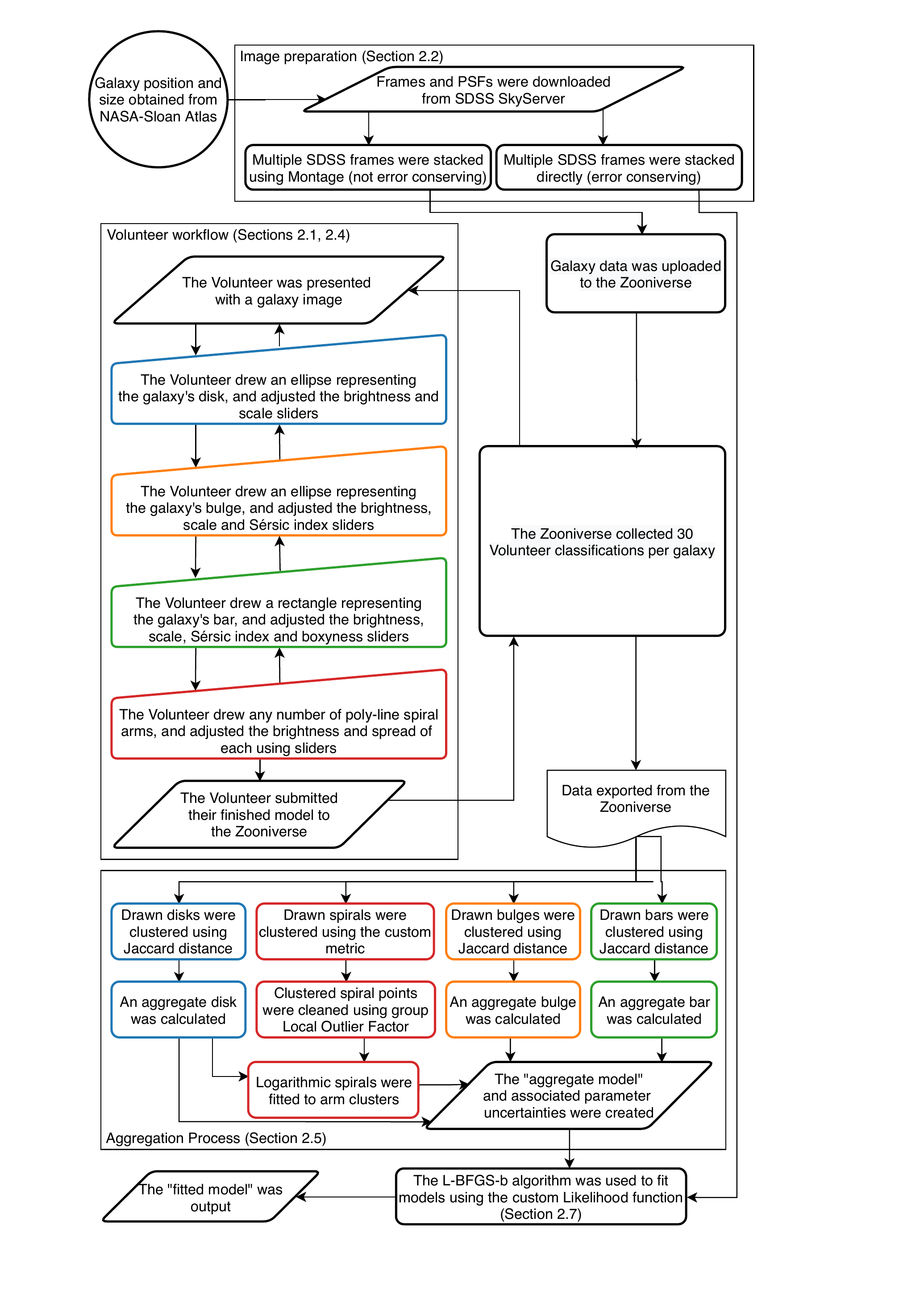}
  \caption{\added{Flowchart detailing the entire \textit{Galaxy Builder} process, from image creation, through classification collection using the Zooniverse, to model aggregation and fitting. Processes, manual input, data inputs and exports, and document exports are displayed distinctly. Colours distinguish between component-specific processes (disc in blue, bulge in orange, bar in green and spiral in red). Black nodes relate to the galaxy as a whole.}}
  \label{fig:gzb-flowchart}
\end{figure*}

\added{This Section describes the \textit{Galaxy Builder} project, and the sample of galaxies and syntetic images used in this paper to examine the project's output. The entire process is summarised in flowchart form in Figure \ref{fig:gzb-flowchart}, with appropriate sections referenced therein.}

\subsection{The \textit{Galaxy Builder} Zooniverse project}

\textit{Galaxy Builder} is a citizen-science project built on the Zooniverse web platform. It asks volunteers to perform detailed photometric modelling of spiral galaxies (potentially including bulge, disc, bar and spiral arm components). A project of this kind, allowing volunteers to interact with and model data, had never been attempted inside the current Zooniverse web platform before, so this project involved designing and implementing a model rendering\footnote{We use the term rendering in a similar manner to that used for computer graphics: to calculate an image from a model or set of rules.} suite inside the existing Zooniverse front-end code-base. As with all citizen science solutions, we had to not only consider the accuracy of the resulting model, but also user experience and engagement in our design decisions.

The closest relative to this project within the Zooniverse ecosystem was the Galaxy Zoo: Mergers project \citep{Holincheck2016:1604.00435v1}. This project asked volunteers to help match the morphological properties of an image of merging galaxies to a plethora of restricted three-body simulations. Galaxy Zoo: Mergers required volunteers to download a Java applet to take part in model selection, while \textit{Galaxy Builder} operates purely inside a web page.

\subsubsection{Project Timeline and Development}

The \textit{Galaxy Builder} project was built inside the Zooniverse's \citep{Simpson:2014:ZOW:2567948.2579215} \textsc{Panoptes-Front-End}\footnote{\url{http://github.com/zooniverse/Panoptes-Front-End}} codebase, using the \textsc{React.js}\footnote{\url{https://reactjs.org/}} framework, as well as WebGL\footnote{\url{https://www.khronos.org/webgl/}} to enable low-latency photometric galaxy model rendering. \textit{Galaxy Builder} entered a Zooniverse beta in late November 2017 and after some user experience improvements and code refactoring, the project was launched as an official Zooniverse project on the 24th of April 2018.

A major challenge during the development of the project was finding the right balance between keeping a simple and intuitive interface and workflow while also allowing the freedom and versatility to properly model galaxies. It was also a significant challenge to develop a compelling and simple tutorial for what is one of the most complex projects attempted on the Zooniverse platform. Feedback from expert users was essential to this process as part of the typical beta trial process for Zooniverse projects\footnote{\url{https://help.zooniverse.org/best-practices/}}.

\subsubsection{The project interface}

\begin{figure}
  \plotone{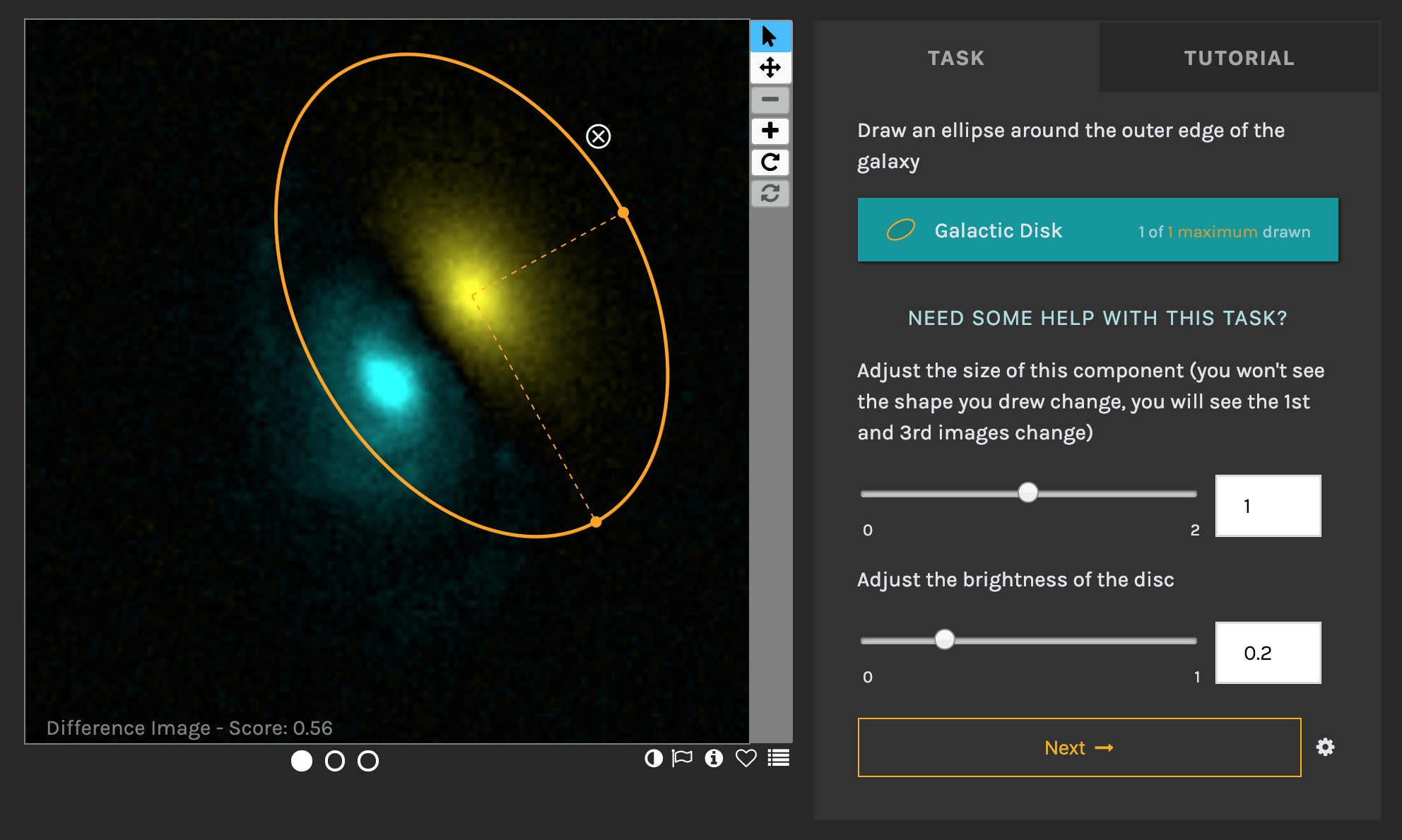}
  \caption{The \textit{Galaxy Builder} interface. The residual image is being shown, and the volunteer is on the ``Disc'' task. The drawn disc component (yellow) is offset from the galaxy image (blue) to demonstrate the positive and negative residuals. Where the image equals the model the residual is black. The dots below the residual image allow the user to switch images. The icons to the right allow panning and zooming of the image (rotation was not functional for this project). The icons to the bottom right of the image allow colour inversion of the galaxy cutout, flagging of the image as inappropriate, inspection of galaxy metadata (i.e. sky position, link to SDSS SkyServer), ability to save the image as a favourite and to add to a Zooniverse ``collection''. The Score shown in the bottom left of the image is calculated using Equation \ref{eqn:gal_score} and is a rough goodness-of-fit measure.}
  \label{fig:interfaceInProgress}
\end{figure}

The \textit{Galaxy Builder} project prompts volunteers to work through the step-by-step creation of a photometric model of a galaxy (described in detail in Section \ref{section:galaxy-model}). A screenshot of the interface can be seen in Figure \ref{fig:interfaceInProgress}, where a residual image is shown. The interface presents a volunteer with three views, which they can switch between at any time: a $r$-band cutout image of a spiral galaxy (see Section \ref{sec:data}), the galaxy model they have created so far, and the residual between their model and image (shown in blue and yellow).

\begin{figure}
  \plotone{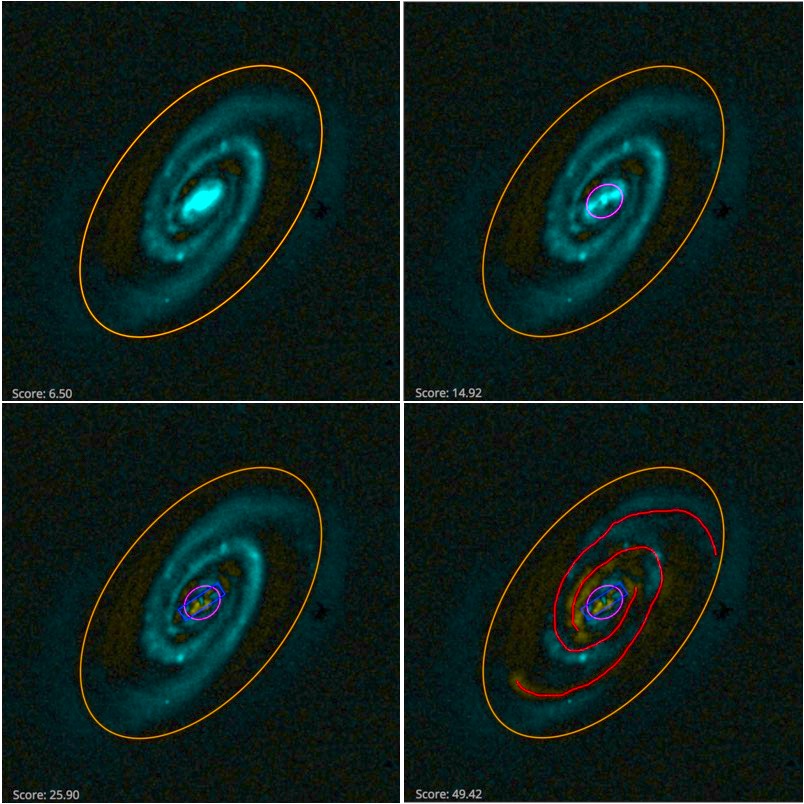}
  \caption{Figure demonstrating the desired result of each step of the modelling process, as seen from the residual image provided to volunteers. The top left panel shows the galaxy after only a disc component has been added: the top right contains a disc and a bulge; the bottom left has a disc, bulge and bar; the bottom right is the finished model with a disc, bulge, bar and spiral arms. The image shown is SDSS J104238.12+235706.8. This brightness and contrast of this image have been edited to improve visibility in print.}
  \label{fig:residualsStepByStep}
\end{figure}

The workflow is designed so that volunteers slowly subtract increasing amounts of light from the galaxy, as is illustrated in Figure \ref{fig:residualsStepByStep}. A tutorial is available that contains a step-by-step guide to completing a classification. At each step, volunteers are asked to first draw a simple isophote, and then make use of a series of sliders to adjust the parameters of the model component (see Section \ref{section:galaxy-model} for more information).

Volunteers are also guided by a ``score'', which is tied to the residuals and chosen to increase from zero to some arbitrary value depending on the galaxy; a less noisy and more easily modelled galaxy will have a higher maximum score. To map a residual image to a final score shown to volunteers we used

\begin{equation}
  \label{eqn:gal_score}
    S = 100 \exp\left(\frac{-A}{N}\sum_{i=0}^N\frac{\text{arcsinh}^2\left(\,|\text{y}_i - M_i|\ /\ 0.6\right)}{\text{arcsinh}\,0.6 }\right),
\end{equation}

where $N$ is the total number of pixels, $y$ is the cutout of the galaxy, normalized to a maximum value of 1 ($y = \text{cutout}/\text{max}(\text{cutout})$), $M$ is the model calculated by volunteers and $A=300$ is an arbitrary choice of scaling chosen based on a handful of test galaxies.

This score has the advantage of being easy (and fast) to generate from the residual image shown to volunteers (which was Arcsinh-scaled in a manner described by \citealt{Lupton2003:astro-ph/0312483v1}), however, it is overly sensitive to small deviations of the model from the galaxy.

\subsection{Sample Selection: Images and Ancillary Data}
\label{sec:data}

As a proposed solution to the problem of fitting multi-component and complex galaxies, \textit{Galaxy Builder} finds a niche with a sample of disc galaxies with spiral features. One such sample is the \textit{stellar mass-complete sample} in \citet{2017MNRAS.472.2263H}, which is a sample of \added{relatively} face-on spiral galaxies ($b/a > 0.4$) with and without bars and selected to be complete across stellar masses $9.45 < \log(M_\star / M_\odot) < 11.05$. The test sample we use for the Galaxy Builder project was therefore selected from the \citet{2017MNRAS.472.2263H} sample of \added{relatively} face-on spiral galaxies.

The morphological information required to select spiral galaxies came from the public data release of Galaxy Zoo 2 (\citealt{Willett2013:1308.3496v2}, hereafter GZ2). Each response to a GZ2 morphology question is allocated a $p$ value ranging from 0 to 1, where 0 indicates no volunteers responded positively to that question and 1 indicates all volunteers who classified that galaxy responded positively (i.e. $p_\text{bar} = 0.5$ would indicate $50\%$ of volunteers said a bar was present in a galaxy). Photometric measurements used for selection came from the NASA-Sloan Atlas (\citealt{2011AJ....142...31B}, hereafter NSA). The \textit{stellar mass complete sample} is constructed using the set of criteria detailed in Table \ref{table:sample_selection}.

\begin{deluxetable*}{cc}
  \tablenum{1}
  \tablecaption{The selection criteria used in \citet{2017MNRAS.472.2263H} to create the \textit{stellar mass-complete sample} of 6222 spiral galaxies.\label{table:sample_selection}}
  \tablewidth{0pt}
  \tablehead{
    \colhead{Description} & \colhead{Value}
  }
  \startdata
    Face-on spiral morphological selection. & GZ2 $p_\text{features} \times p_\text{not edge on} \times p_\text{spiral} \ge 0.5$ \\
    Redshift limits. & $0.02 < z < 0.055$ \\
    \replaced{Face-on}{Relatively face-on} galaxy selection using $g$-band axial ratio. & $(b/a)_g > 0.4$ \\
    Mass limits for rough volume limited sample. & $9.45 < \log(M_* / M_\odot) \le 11.05$ \\
    Mass limits for complete sample\tablenotemark{a} & $2.07\log(z) + 12.64 < \log({M_* / M_\odot}) < 2.45\log(z) + 14.05$ \\
  \enddata
  \tablenotetext{a}{Stellar masses from \citet{2014ApJS..210....3M}}
  \tablenotetext{a}{Using the method of \citet{Pozzetti2009:0907.5416v2} and limits calculated by \citet{2017MNRAS.472.2263H}}
\end{deluxetable*}

\begin{figure*}
  \plotone{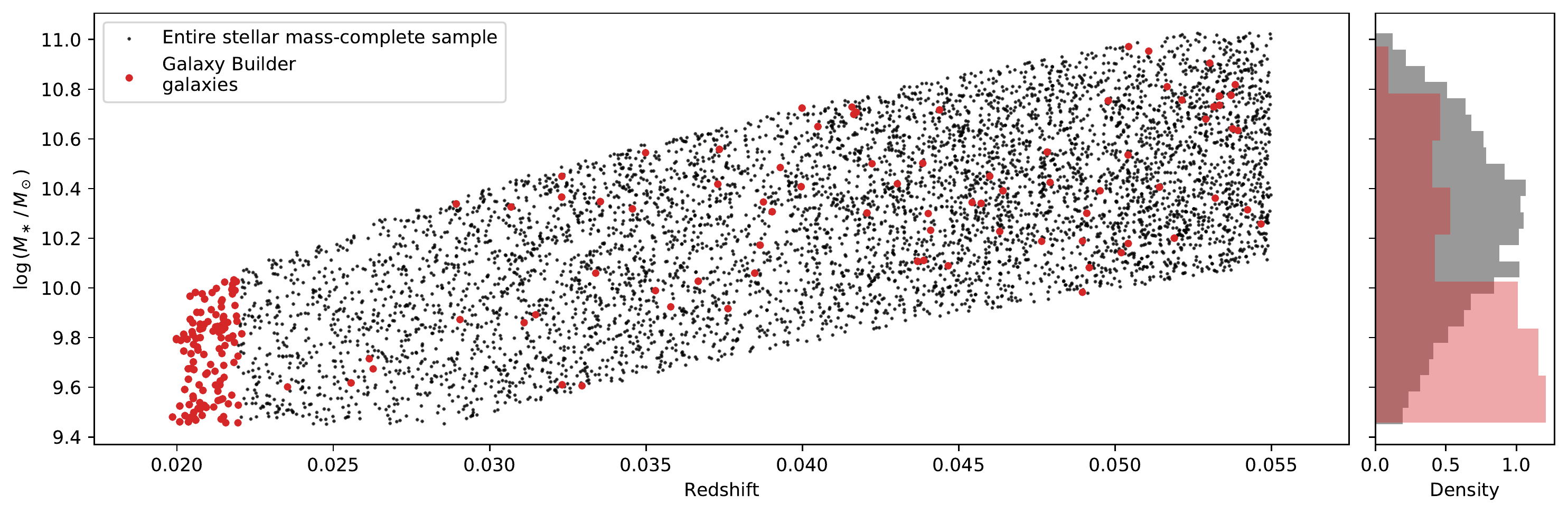}
  \caption{Redshift against total galaxy stellar mass for all galaxies in the \textit{stellar mass-complete sample}, with the 198 galaxies considered in this paper highlighted in red. The distribution of stellar masses is shown in the right panel for the total sample and for the galaxies considered here. It is evident that the galaxies for which we collected classifications are not complete in stellar mass, but it is possible to select a further subset that would be.}
  \label{fig:mass_redshift}
\end{figure*}

The \textit{stellar mass-complete sample} was split into smaller sub-samples, each containing 100 galaxies. In an iterative process, each sub-sample was chosen to contain the 60 lowest redshift unclassified galaxies, and 40 random unclassified galaxies. This was done to ensure we would have an early sample to work with given the {\it a priori} unknown rate at which volunteers would provide classifications. Due to time constraints, classifications were only collected for two unique sub-samples. The mass-redshift relation of galaxies in the \textit{stellar mass-complete sample} from \citet{2017MNRAS.472.2263H} can be seen in Figure \ref{fig:mass_redshift}, with galaxies present in this work highlighted in red. Stellar Masses were calculated by \citet{2014ApJS..210....3M}.

In the first two sets of 100 galaxies, 1\% of galaxies (i.e. 2 images) failed to run through the image preparation process, due to an error when attempting to montage multiple frames. The root cause of this error is unknown, but it leaves a sample of 198 galaxies with images (the \textit{test sample}, 98 of which are repeated in a validation subset) that are considered in this paper, in order to explain the method used and test the reliability of the models obtained.

\subsubsection{Image and modelling metadata extraction}
\label{sec:image_creation}

The galaxy data shown to volunteers in the \textit{Galaxy Builder} project came in two forms: A grey-scale image cutout of the galaxy and a JSON file containing rendering information for the web-interface.

Both forms of data were obtained using a similar process:

\begin{enumerate}
\item A montage of multiple $r$-band corrected frames from the SDSS DR13 \citep{2017ApJS..233...25A} data release was created. To combine multiple FITS images, we made use of Astropy \citep{2018AJ....156..123A}, and the \textsc{Montage} \citep{2010arXiv1005.4454J} software package.
\item This montage was cropped to four times the Petrosian radius of the galaxy.
\item The \textsc{SExtractor} software \citep{source-extractor} was used to identify regions containing secondary sources (foreground stats, other galaxies) and generate a mask.
\item A PSF was obtained from the relevant Sloan $r$-band \texttt{psField} file, extracted at the central position of the galaxy \citep{2002AJ....123..485S}.
\item The JSON file was written containing the cut-out data and the 2D boolean mask obtained from the source extraction process. This file also contained other metadata needed for the rendering process (PSF, the size of the PSF array, and the width and height of the image).
\item Another JSON file containing simply the information used to render the volunteer's model (image size and PSF) was created.
\item An arcsinh-stretch was applied to the masked cutout (as described by \citealt{Lupton2003:astro-ph/0312483v1}). It was then saved as a grey-scale image.
\end{enumerate}

The decision to use $r$-band images in our subject set was due to its higher signal-to-noise than other bands.

Once a sub-sample had been created, the Zooniverse's \textsc{panoptes-python-client}\footnote{\url{https://github.com/zooniverse/panoptes-python-client}} was used to upload them as a subject-set to the Zooniverse.

The reprojection performed by \textsc{Montage} has a smoothing effect on the data, and thus does not conserve errors. We, therefore, create a separate stacked image, sigma image and corresponding pixel mask, using the same $r$-band corrected frames present in the montage. These images were not shown to volunteers but were used for model fitting and comparison.

\subsection{Choice of Retirement limit}
\label{sec:retirement-limit}

The number of independent answers needed to create reliable and reproducible aggregate classifications was not known at the start of this project. An initial experiment with collecting 10 classifications per galaxy demonstrated that this was insufficient; further experimentation with a diverse range of galaxy types (most with prominent spiral features including grand-design and flocculent arms) revealed 30 classifications per galaxy was sufficient.

The entire {\it test sample} of 198 galaxies was then presented to users, with 30 classifications collected per galaxy. In addition, one of the subsets was presented a second time, thus providing a validation subset to measure consistency between sets of 30 classifications on the same galaxies.

\begin{figure*}
  \plotone{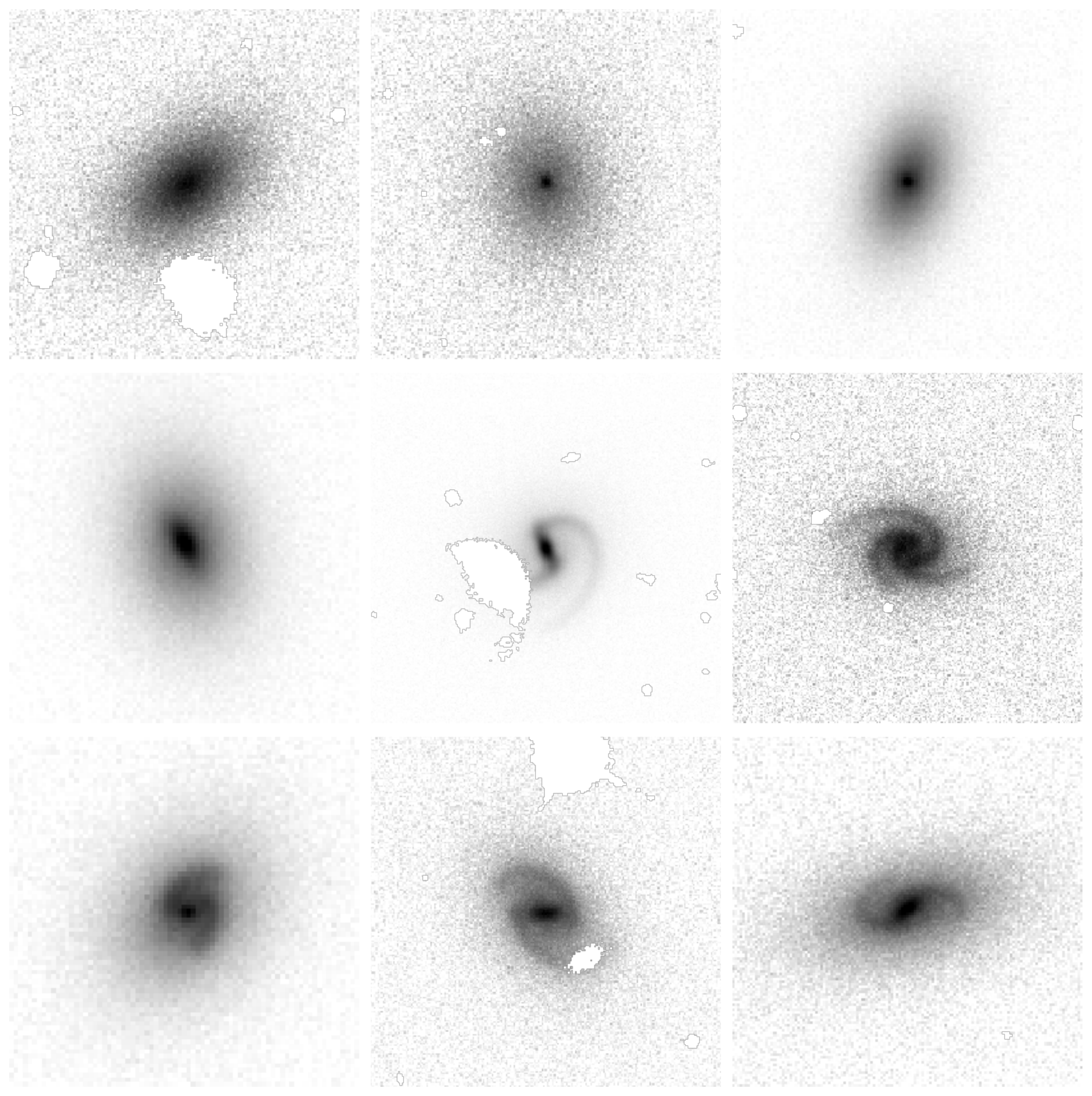}
  \caption{Arcsinh-stretched images of the synthetic galaxies present in the \textit{calibration subset}. These galaxies were designed to look as realistic as possible, while being described perfectly by the model available to volunteers.}
  \label{fig:calibration-subset}
\end{figure*}

We also created 9 synthetic images of galaxies, containing various combinations of components available to volunteers and a spread of possible parameters. These synthetic galaxies were based off of a set of target galaxies from \textit{Galaxy Builder} and designed to be as realistic as possible, including the addition of realistic noise and pixel masks. This set of synthetic images is shown in Figure \ref{fig:calibration-subset} and was used to calibrate our aggregation and fitting methodology and thus is referred to as the \textit{calibration subset}.

\subsection{The Galaxy Model}
\label{section:galaxy-model}

Our chosen galaxy model was largely based on components described in \citet{galfit-paper}. The modelling code ignores masked regions identified as secondary sources by \textsc{SExtractor}. It over-samples the bulge, disc and bar components by a factor of five and performs PSF convolution using a PSF obtained from the relevant Sloan $r$-band \texttt{psField} file, extracted at the central position of the galaxy \citep{2002AJ....123..485S}. The model created by a volunteer could be chosen from

\begin{enumerate}
\item One exponential, ellipsoidal disc.
\item One ellipsoidal S\'ersic bulge, with $n$ chosen by volunteers.
\item One S\'ersic bar with a ``boxiness'' modifier (as described in \citealt{galfit-paper}), with $n$ and $c$ chosen by volunteers.
\item Any number of freehand poly-line\footnote{a poly-line, or polygonal chain, is a series of connected line segments.} spiral arms, as described below.
\end{enumerate}

\subsubsection{Spiral Arm Model}
Each spiral arm is modelled using a poly-line drawn by the volunteer. The brightness of a spiral arm at any point is given by the value of a Gaussian centred at the nearest point on any drawn poly-line, with volunteers able to choose the Gaussian width and peak brightness using sliders. Radial falloff was added by multiplying by the value of the previously added exponential disc, though volunteers could change the half-light radius of this falloff disc.

\subsection{Classification Aggregation Methodology}

\begin{figure*}
  \plotone{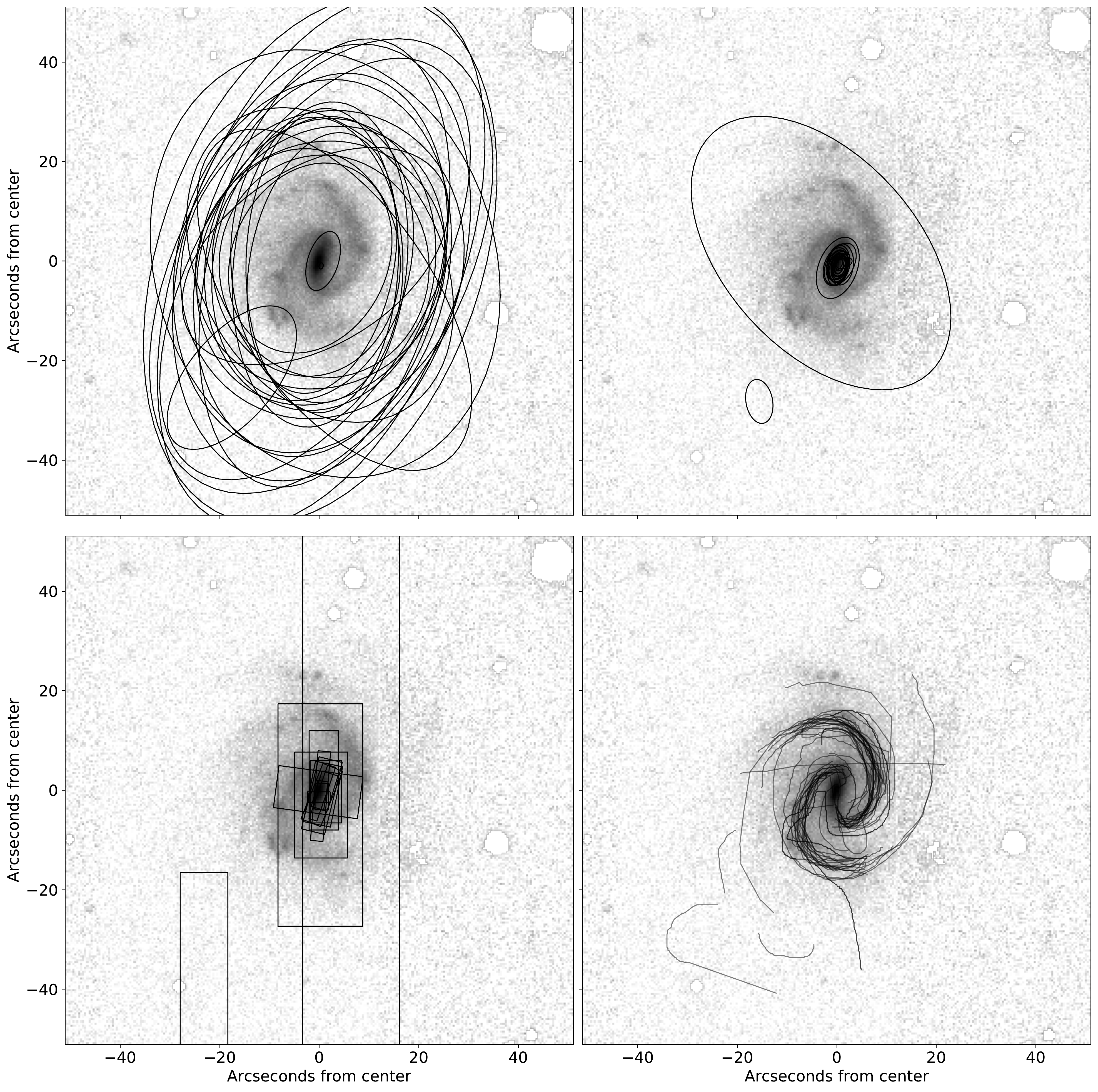}
  \caption{Components drawn by volunteers for UGC 4721. The top left panel shows drawn discs, top right shows drawn bulges, bottom left shows drawn bars and bottom right shows drawn spiral arms. Discs, bulges and bars are displayed at twice their effective radii. These raw marks are subsequently aggregated to produce a consensus value for each galaxy component.}
  \label{fig:drawn_shapes}
\end{figure*}

In this Section, we will use the galaxy UGC 4721, a two-armed barred spiral galaxy at $z=0.02086$ classified by \citet{deVaucouleurs1991} as SBcd, to illustrate the data reduction and aggregation methodology. For UGC 4721 we received 32 classifications, containing 28 discs, 24 bulges, 17 bars and 47 drawn spiral arm poly-lines (four classifications did not contain spirals, seven contained one spiral arm, fourteen contained two arms, six contained three arms and one contained four arms). These annotations can be seen in Figure \ref{fig:drawn_shapes}, overlaid on the greyscale $r$-band image of the galaxy.

\subsubsection{Aggregation of Volunteer Models}
\label{sec:aggregation_of_volunteer_models}

Aggregate model calculation was done on a component-by-component basis, rather than per classification, i.e. clustering of discs was performed independently to that of bulges, bars and spirals. We did not take into account any slider values, only the shape drawn by the volunteers. Disk classifications were doubled in effective radius to correct for a systematic error in disk size observed in the classifications received for the \textit{calibration subset}. Model parameters were restricted to be within the limits shown in Table \ref{table:bad_values} (deemed to be the physically acceptable bounds). All components were transformed from the coordinate space of the \textsc{Montage}-created images to the more accurate stacked images created for fitting. Clustering was performed using the Jaccard distance measure (also known as the intersect-over-union distance, or IOU distance), which is a simple metric determining the relative shared area of two sets:

\begin{equation}
d_J(A, B) = 1 - \frac{|A \cap B|}{|A \cup B|}.
\end{equation}

The algorithm chosen to perform clustering was the \replaced{density-based spatial clustering of applications with noise}{Density-Based Spatial Clustering of Applications with Noise} (DBSCAN, \citealt{dbscan}) algorithm, due to its robustness and speed. We made use of Scikit-learn \citep{scikit-learn} to implement the algorithm. In DBSCAN the core of a cluster is defined as a group of at least \replaced{\texttt{min\_points}}{$N_\mathrm{min}$} items that are all within a distance \replaced{\texttt{eps}}{$\epsilon$} of each other. Additionally, any points within a distance \replaced{\texttt{eps}}{$\epsilon$} of a cluster's core are also associated with the cluster.

\begin{deluxetable*}{rlcc}
  \tablenum{2}
  \tablecaption{\added{The maximum, minimum and default values for model parameters. Model parameters are defined in Appendix \ref{sec:appendix-model-fitting}. Note that some parameters were allowed to overflow when fitting, for instance an axis ratio greater than 1 (signifying a swap of major and minor axis) was allowed, and corrected for once fitting reached completion. This helped avoid the optimizer encountering parameter bounds and failing to converge. Component position angle ($\psi$) and spiral pitch angle ($\phi$) were similarly unconstrained.\label{table:bad_values}}}
  \tablewidth{0pt}
  \tablehead{
    \colhead{Component} & \colhead{Parameter} & \colhead{Tuning Minimum Bound} & \colhead{Tuning Maximum Bound}
  }
  \startdata
    disc      & $\mu_x$    & -inf           & inf            \\
              & $\mu_y$    & -inf           & inf            \\
              & $\psi$     & -inf           & inf            \\
              & $q$        & 0.25           & 1.2            \\
              & $R_e$      & 0              & inf            \\
              & $I_e$      & 0              & inf            \\
    bulge     & $\mu_x$    & -inf           & inf            \\
              & $\mu_y$    & -inf           & inf            \\
              & $\psi$     & -inf           & inf            \\
              & $q$        & 0.6            & 1.2            \\
              & $R_e\ /\ R_{e,\,\mathrm{disc}}$ & 0.01 & 1   \\
              & $(B/T)_r)$ & 0              & 0.99           \\
              & $n$        & 0.5            & 5              \\
    bar       & $\mu_x$    & -inf           & inf            \\
              & $\mu_y$    & -inf           & inf            \\
              & $\psi$     & -inf           & inf            \\
              & $q$        & 0.05           & 0.5            \\
              & $R_e\ /\ R_{e,\,\mathrm{disc}}$ & 0.05 & 1   \\
              & $(B/T)_r)$ & 0              & 0.99           \\
              & $n$        & 0.3            & 5              \\
              & $c$        & 1              & 6              \\
    spiral    & $I_s$      & 0              & inf            \\
              & $A$        & 0              & inf            \\
              & spread     & 0              & inf            \\
              & $\phi$     & -85            & 85             \\
              & $\theta_\mathrm{min} $ & -inf & inf          \\
              & $\theta_\mathrm{max} $ & -inf & inf          \\
  \enddata
\end{deluxetable*}

\subsubsection{Disc, Bulge and Bar Clustering}

We select the disc clustering hyperparameters such that a disc is clustered for all galaxies, and the bulge hyperparameters to most successfully recover the morphology of galaxies in the \textit{calibration subset}. The value of \replaced{\texttt{eps}}{$\epsilon$} used to cluster bars was tuned such that the aggregate model best agreed with GZ2 $p_\mathrm{bar}$ ($p_\mathrm{bar} < 0.2$ implying no bar and $p_\mathrm{bar} > 0.5$ implying a definite bar). The values chosen for \replaced{\texttt{eps}}{$\epsilon$} were 0.3, 0.4, 0.478 for the disc, bulge and bar; \replaced{\texttt{min\_points}}{$N_\mathrm{min}$} was set to 4 for all three of these components.

We define the aggregate component to be the shape that minimises the sum of Jaccard distances to each of the members of the cluster. For our example galaxy, UGC 4721, clustered and aggregate components can be seen in Figure \ref{fig:mean_shapes}.

\subsubsection{Spiral Arm Clustering}
\label{sec:spiral_clustering}
To cluster drawn spiral arms, we define a custom separation measure to represent how far away one poly-line is from another. This measure was chosen to be the mean of the squared distances from each vertex in a poly-line to the nearest point (vertex or edge) of another poly-line, added to the mean of the squared distances from the second poly-line to the first. We make use of this separation measure inside the DBSCAN algorithm to cluster these drawn lines, after removing any self-intersecting drawn arms (as this was deemed an easy method to filter out ``bad'' classifications). Values of 0.001 and 4 were used for the \replaced{\texttt{eps}}{$\epsilon$} and \texttt{min\_samples} hyper-parameters respectively.

Once spiral classifications on a galaxy have been clustered into the physical arms they represent, the points are deprojected using the axis ratio and position angle of the aggregated disc. The deprojection method assumes a thin disc and stretches the ellipsoidal minor axis to match the major axis.

\begin{figure*}
  \plotone{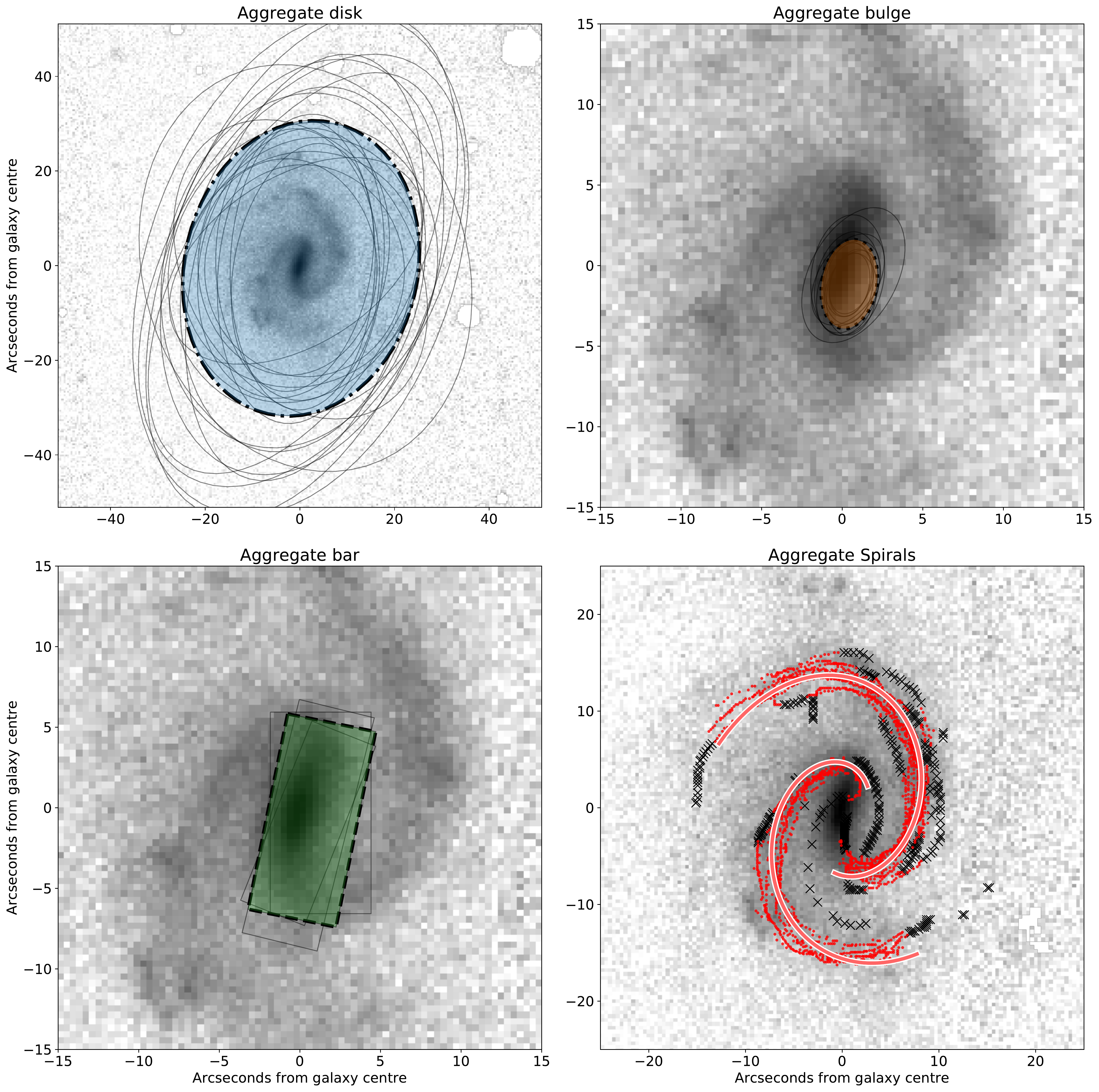}
  \caption{Calculated aggregate components for UGC 4721. The aggregate disc is shown using a dot-dashed line and blue fill in the upper left panel, the aggregate bulge with a dotted line and orange fill in the upper right panel, the aggregate bar using a dashed line and green fill in the lower-left panel and the aggregate spiral arms are plotted as red lines in the lower right panel. S\'ersic components are displayed at twice their effective radii. Black crosses in the lower right panel indicate spiral arm points that were identified as outliers and removed during cleaning (described in Section \ref{sec:spiral_clustering}). The aggregated components agree well with the underlying morphology, despite the noisiness of the classifications received.}
  \label{fig:mean_shapes}
\end{figure*}

Deprojected points within each drawn poly-line are converted to polar coordinates and unwound to allow model fitting. These unwound points are then cleaned using the Local-outlier-factor algorithm (LOF, \citealt{local-outlier-factor}). For each drawn poly-line in the cluster, the LOF algorithm was trained on all points not in that arm, and then used to predict whether each point in the arm should be considered an outlier. In this way, we clean our data while respecting its grouped nature. The points removed as outliers for the example galaxy are shown in the bottom right panel of Figure \ref{fig:mean_shapes}.

For each arm cluster in each galaxy, a logarithmic spiral model was fitted using Bayesian Ridge Regression, performed using the Scikit-learn python package. A logarithmic spiral was chosen due to its simple form with a constant pitch angle. Hyperpriors on the noise parameter were chosen by fitting a truncated gamma distribution \citep{2014arXiv1401.0287Z} to the spiral width slider values returned by volunteers (ignoring sliders left at the default or moved to the extremes of allowed values). Any logarithmic spirals within a distance of 0.0005 (given by the clustering metric) were deemed to be from the same arm and thus their classifications were merged and a log-spiral recalculated.

We do not assume that every arm in a galaxy has the same pitch angle. To obtain a single value for the pitch angle of a galaxy, we take the length-weighted average pitch angle of all arms detected in the galaxy (as used by \citealt{Davis2014:1402.1910v1}).

The galaxy model for UGC 4721 obtained through aggregation can be seen in the bottom left panel of Figure \ref{fig:model_tuning}.

\subsection{Error Estimation of Aggregate models}
\label{sec:error_estimation}

As all components in a cluster can be viewed as volunteers' attempts at modelling the true underlying component, the sample variance of the parameters of these shapes can be used as a measure of confidence in the parameters present in the aggregate result. These are highly sensitive to clustering hyperparameters, and are only valid for a component's position, size and shape. Figure \ref{fig:mean_shapes} illustrates the variance in clustered shapes for our example galaxy (UCG 4721); we see a large variation in the clustered discs, and much closer agreement on the bulge and bar size and shape.

\subsection{Model Fitting}

The final step in creating \textit{Galaxy Builder} models is a numerical fit to fine-tune parameters. This fitting was performed using the L-BFGS-b algorithm \citep{doi:10.1137/0916069}, implemented in \textsc{Scipy} \citep{scipy-paper}. We minimize a custom likelihood function that assumes Gaussian error on pixel values and incorporates the priors on parameters we obtain from clustering. The full fitting model and likelihood is detailed in Appendix \ref{sec:appendix-model-fitting}. We use the same model as used by volunteers in the online interface (with altered limits), with spiral arms restricted to being logarithmic spirals relative to the disc, and without the ability to change the relative falloff of spiral arms.

The model rendering and fitting code was written up using Google's JAX package \citep{jax2018github}, which allows GPU-optimization and automatic gradient calculation, enabling quick and accurate calculation of the jacobian matrix needed for the L-BFGS-b minimization algorithm.

\begin{figure*}
  \plotone{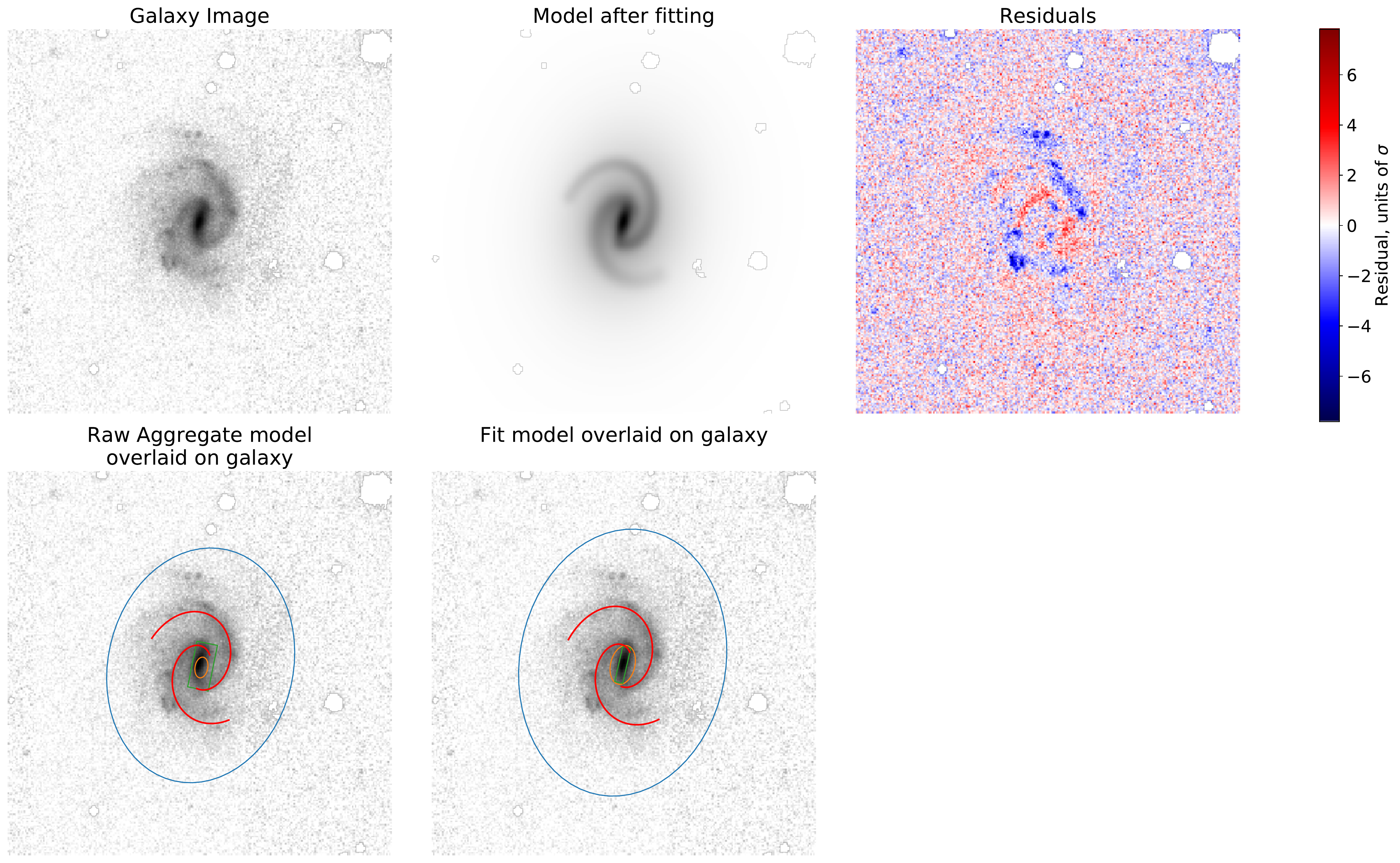}
  \caption{Effect of fitting on the aggregated models. The top left panel shows an Arcsinh-scaled image of the galaxy being fit (UGC 4721), the top middle shows the final model obtained (with the same limits and scaling as the galaxy image) and the top right shows the difference between the two images, in units of pixel uncertainty. The bottom panels show a simple representation of the model before and after tuning, overlaid on the galaxy image from the top-left panel. With minimal change to the aggregated components, we recover a detailed model that matches the galaxy exceptionally well, as evident in the residuals.}
  \label{fig:model_tuning}
\end{figure*}

We initially fit only for the brightnesses of components, and then simultaneously for all free parameters of all components. The result of the fit, including the final photometric model for UGC 4721, can be seen in \ref{fig:model_tuning}. The secondary components have been accounted for well, and the model has a sensible reduced chi-squared value of 1.176, where we have approximated degrees of freedom as the number of unmasked pixels present in the galaxy image (similar to \textsc{Galfit}).

We use the errors described in Section \ref{sec:error_estimation} as parameter uncertainties, as we feel an approach based on the local curvature of the likelihood-space (as used by \texttt{Galfit}) would likely fall foul of the issues described in the introduction and thus be an under-estimate. This decision means we do not have uncertainties for some parameters.

We remove two models for which a fit did not converge.

\section{Results}
\label{sec:results}

In this Section we present \textit{Galaxy Builder} models for 198 galaxies, from the aggregation of user classifications (aggregate models), and with parameters fine-tuned by a numerical fit (fitted models). We explore the consistency with which volunteers modelled galaxies, the accuracy of the aggregate models, and compare the aggregate and fitted models to comparable results in the literature.

\subsection{The Calibration Set}
\label{sec:calibration-set-results}

The calibration subset was a set of nine synthetic galaxy images created from \textit{Galaxy Builder} models, which were then re-run through the \textit{Galaxy Builder} process. These galaxies were used to fine-tune clustering and fitting hyperparameters (See Section \ref{sec:aggregation_of_volunteer_models}), as the ground truth was known. Our ability to recover morphology accurately is essential validation for our ability to recover good photometric models of galaxies.

\begin{figure*}
  \plotone{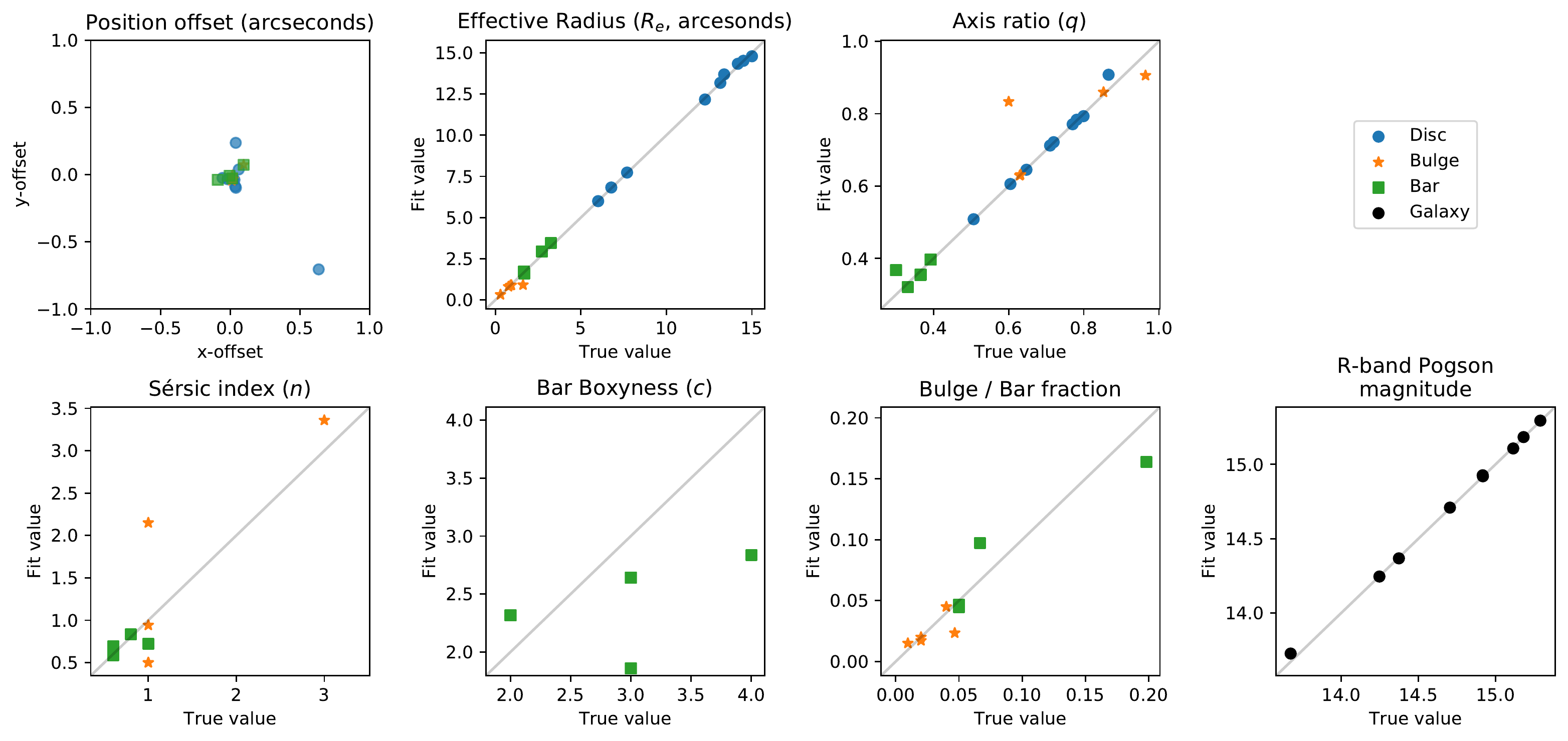}
  \caption{Plots examining the accuracy of fit parameters for the calibration subset of galaxies. Most parameters are recovered to a high degree of accuracy, however S\'ersic index and boxiness are difficult to determine only using gradient descent, \added{as they do not significantly impact the goodness of fit \citep{2012MNRAS.421.2277L}}. The error in the fit values reflects this problem.}
  \label{fig:calibration_parameter_recovery}
\end{figure*}

\added{The scatter between true and measured parameters is shown in Figure \ref{fig:calibration_parameter_recovery};} these results highlight the importance of good priors to obtain accurate fits of complex photometric models. In more detail, the models recovered for the nine synthetic galaxy images demonstrate that:

\begin{enumerate}
  \item Model parameters were generally recovered to a high degree of accuracy
  \item We successfully recover all spiral arms present, and do not receive any false positives
The spiral pitch angles obtained through aggregation vary by $< 9\degree$ from the true values, with fitting improving this error slightly.
  \item Volunteers systematically use elongated bulges to model bar components. This resulted in two false positives for bulge presence in the aggregate models. This feature (switching light between model components) is a common issue in all photometric fitting methods \citep{2018MNRAS.473.4731K}.
  \item The Jaccard metric is unstable to small changes in rotation for highly elliptical components (i.e. bars). This resulted in one false negative of bar presence in the aggregate model.
\end{enumerate}

The fitting step for this subset of images highlighted the benefit of obtaining a rough starting point through clustering of user classifications; the method struggled to recover structural parameters for which we did not obtain such a starting point (S\'ersic index and bar boxiness). These parameters are difficult to identify using gradient descent \citep{2012MNRAS.421.2277L}, suggesting future work should attempt to obtain priors on these parameters from volunteers and make use of a more robust fitting algorithm.

\subsection{Examination of Volunteer consistency}
We aggregate two independent models for a set of 98 galaxies based on ``original'' or repeat (``validation'') classifications, obtained with the same retirement limit (see Section \ref{sec:data} for more on this selection).

\begin{figure*}
  \plotone{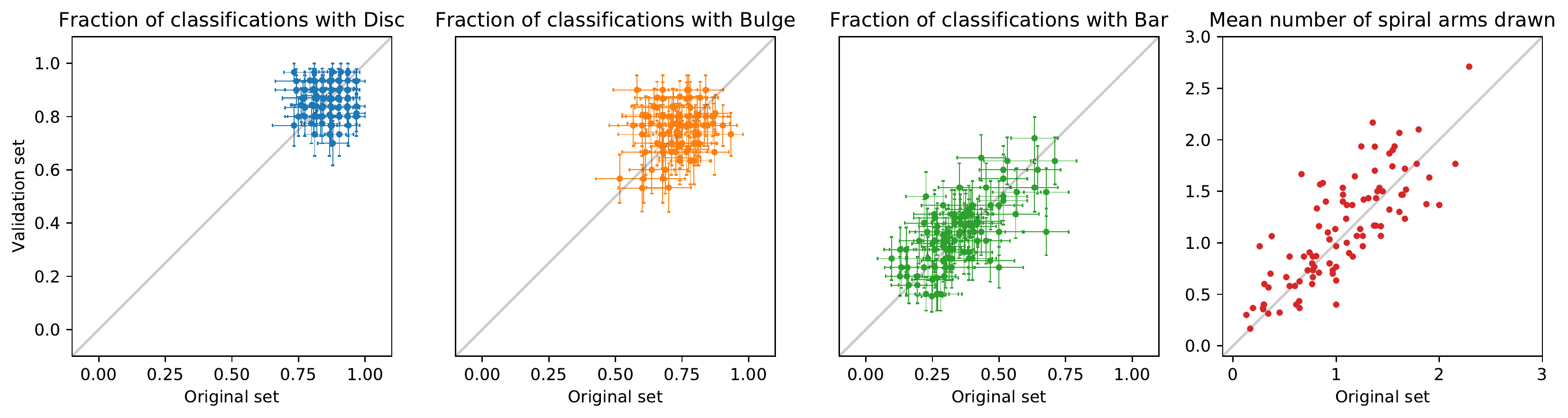}
  \caption{Comparison of frequency of use of component in volunteer models between the original and validation sets of classifications. Errors shown on the disc, bulge and bar arise from Binomial error estimation. We see that classifications are generally consistent within errors, validating our assumption of volunteer independence.}
  \label{fig:volunteer_component_consistency}
\end{figure*}

One of the simplest choices the volunteers have is whether to include a model component or not. Figure \ref{fig:volunteer_component_consistency} illustrates the consistency with which volunteers made use of a component in their model for a galaxy. We see that volunteer classification is very consistent, with scatter as predicted by the Binomial uncertainty on the mean. Volunteers almost always make us of a disc and bulge (as seen in the \textit{calibration subset}), and bulge, bar and spiral arm usage is consistent within Binomial error. One common challenge is that some volunteers used a very ellipsoidal bulge and the ends of spiral arms to model light that other users modelled with a bar. This caused some scatter in aggregate models.

In the end, the aggregated validation model is identical to the original aggregated model in around 40\% of galaxies. The most common changes are a missing bar component or a missing single spiral arm. This may suggest that more than 30 classifications should be collected per galaxy, or could be an artefact of the lack of consensus among volunteers for galaxies with difficult-to-determine components.

\begin{figure*}
  \plotone{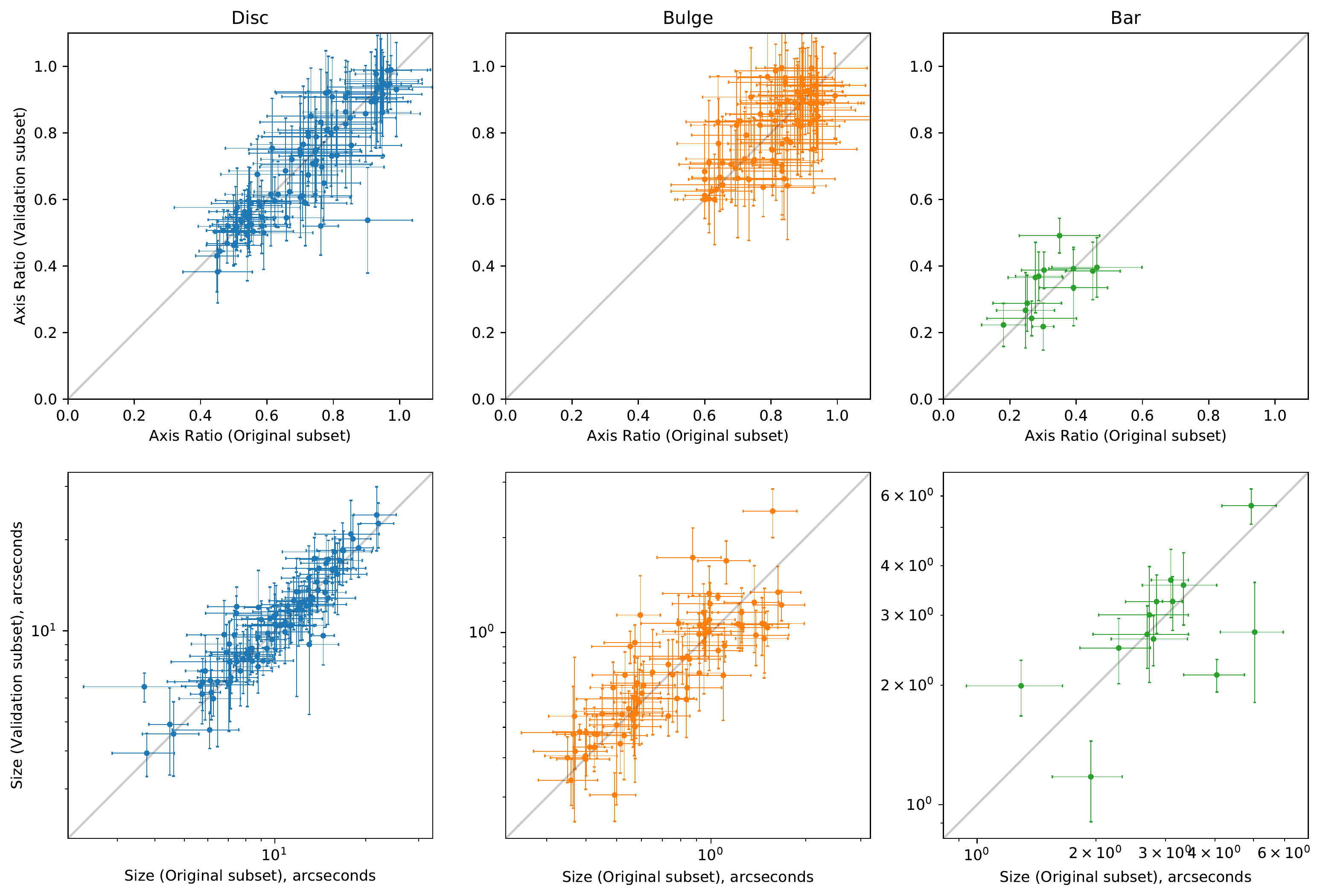}
  \caption{Comparison of component shape in aggregate models between the original and validation sets. Errors are obtained through the sample variance of clustered components, as detailed in Section \ref{sec:error_estimation}. We see close agreement between aggregate components from the two sets, suggesting that the clustering method is robust to the scatter in classifications.}
  \label{fig:aggregate_model_consistency}
\end{figure*}

After selecting a component, the volunteer sets its shape and size. The variation in axial ratios and effective radii for the aggregate discs, bulges and bars are shown in Figure \ref{fig:aggregate_model_consistency}. The aggregate discs and bulges are consistent within errors, however, bars show more scatter. Bars are one of the most challenging components to aggregate consistently. This is partly because even a strongly barred galaxy with 30 classifications overall might receive only 15 or so drawn bars, and lower numbers of classifications result in more scatter. In addition, the aggregation method is more sensitive to rotation of highly elongated shapes. Both factors probably contribute to lower consistency in bar components.

\subsection{Comparison to results in the literature}

After having aggregated and fitted models for our galaxies, we examine how our models compare to other results in the literature. Part of the motivation for exploring the \textit{Galaxy Builder} method was that there exists no published large sample of galaxies with four-component photometric fits. This means we can only make comparisons for individual or subsets of model components (e.g. just disc and bulge) and by design \textit{Galaxy Builder} models will differ as we have attempted to fit bulge-disc-bar-spiral models to all our galaxies. The reader is therefore cautioned against treating literature models as any kind of ``ground truth'' since deviation from these simple models is part of the goal of this project. We provide these comparisons not to check how well our models work, but to provide data on how they compare with other well known, but much simpler photometric models.

\subsubsection{Comparison to Galaxy Zoo morphology}

The simplest comparison we can make to external results is to examine whether our models respect the existing morphological classifications present in the literature. We make use of Galaxy Zoo 2 (\citealt{Willett2013:1308.3496v2}, hereafter GZ2) results, including the redshift debiasing described in \citet{Hart2016:1607.01019v1} and spiral properties calculated in \citet{Hart2016:1607.01019v1}.

When comparing the probability of a volunteer's classification containing a bar component against a galaxy being classed as strongly-barred or as having no bar (as defined in \citealt{Masters2010:1003.0449v2}), we see reasonable agreement. Classifications of GZ2 strongly-barred galaxies ($p_\text{bar} > 0.5$) are more likely to contain a bar than GZ2 unbarred galaxies ($0.47 \pm 0.15$ vs. $0.29 \pm 0.11$). While there is some overlap in these probabilities, the Pearson correlation between GZ2's $p_\text{bar}$ and the bar likelihood in \textit{Galaxy Builder} is $0.56$, implying a significant correlation. We also note that GZ2 bar classifications exclude most weak bars \citep{2017MNRAS.469.3363K}.

\begin{figure}
  \plotone{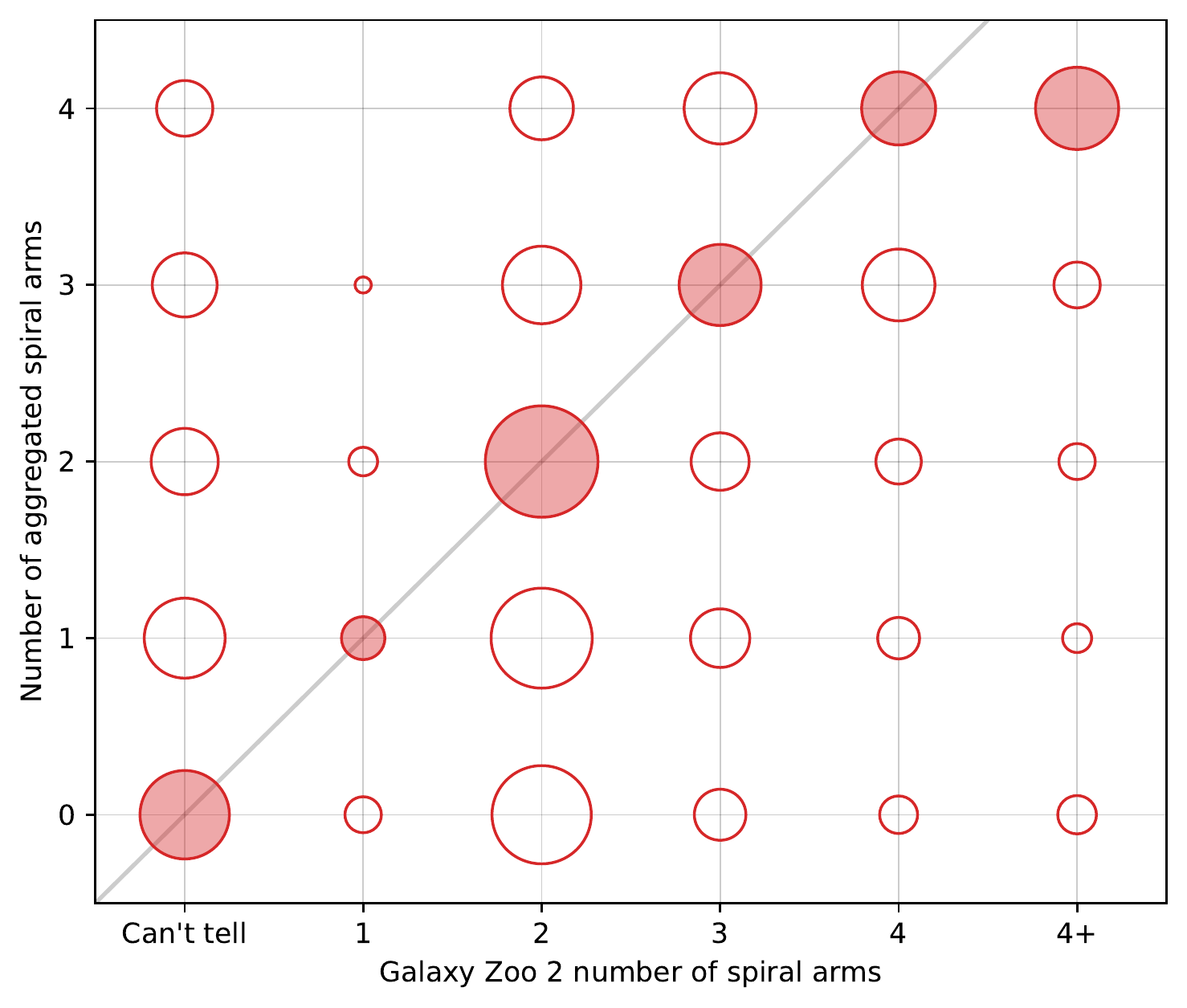}
  \caption{Density plot of GZ2 vote counts for spiral arm number vs the number of spiral arms obtained through aggregation. The area of each circle can be seen as the level of agreement between \textit{Galaxy Builder} aggregate models and GZ2 classifiers, and is defined by Equation \ref{eq:spiral_circle_area_size}. The circle with the largest area for each possible GZ2 response is highlighted by shading. The 1:1 relationship suggests the clustering method is correctly recovering the behaviour of volunteers.}
  \label{fig:n_spirals_comparison}
\end{figure}

We also compare the number of spiral arms aggregated by \textit{Galaxy Builder} with the responses to the GZ2 ``number of arms'' question (of which the possible responses were one, two, three, four, more than four or ``Can't tell''). We attempt to account for the spread in volunteer answers to this question by binning responses, rather than using the mean or modal response. The results of this comparison can be seen in Figure \ref{fig:n_spirals_comparison}. The area of each circle can be seen as the level of agreement between \textit{Galaxy Builder} aggregate models and GZ2 classifiers, it is defined as

\begin{equation}
  \label{eq:spiral_circle_area_size}
  A_{i, j} \propto \sum_{k}^{N_g}\frac{1}{M_k}\sum_{m}^{M_k}
  \begin{cases}
    1,&\ \mathrm{if}\ n_k = i\ \mathrm{and}\ C_{k, m} = j\\
    0,&\ \mathrm{otherwise}
  \end{cases},
\end{equation}

where $n_k$ is the number of aggregate arms for galaxy $k$ (out of $N_g$ galaxies), $C_{k, m}$ is the $m$-th answer for galaxy $k$ (out of $M_k$ answers).

The circle with the largest area for each possible GZ2 response is highlighted, and agrees with the number of spiral arms aggregated here for $m=1, 2, 3, 4$. No aggregate model contained more than four spiral arms, and when galaxies have an uncertain number of spiral arms (the ``Can't tell'' GZ2 response) we mostly do not aggregate any spiral arms.

It is not uncommon in \textit{Galaxy Builder} for one spiral arm to have been broken into two smaller segments. We also occasionally identify two distinct clusters that represent the same physical arm. These two reasons account for a majority of cases where GZ2 classifications suggest a galaxy has two spiral arms and we have clustered a larger number. Improved project user experience would be crucial in correcting these errors.

\subsubsection{Comparison to One-component fit - axis ratio}

\begin{figure}
  \plotone{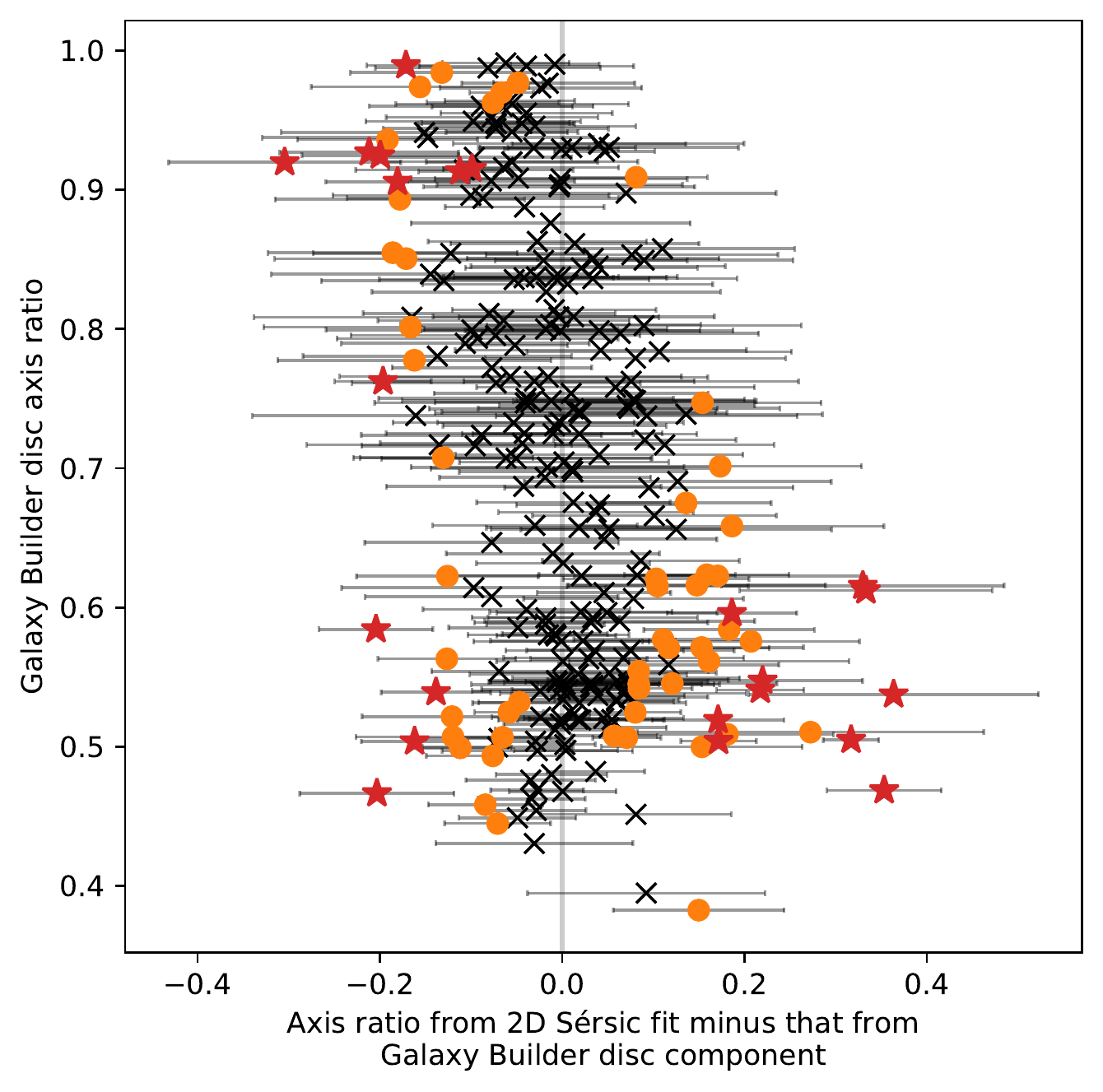}
  \caption{Difference between the axis ratios of the aggregated disc component (before fitting) to the results of an $r$-band S\'ersic profile fit. Points  between one- and two-sigma are highlighted as orange squares, points outside $2\sigma$ are shown as red stars. While the overall relationship is good, the increase prevalence of points outside $2\sigma$ is a clear indication of bias caused by the \textit{Galaxy Builder} online user interface.}
  \label{fig:ax_ratio_comparison}
\end{figure}

We compare the axis ratios of the discs of \textit{Galaxy Builder} aggregate models (without fitting) to the axis ratio of a 2D S\'ersic fit to the $r$-band SDSS image of each galaxy (as provided in the NSA catalog, \citealt{2011AJ....142...31B}). \added{The resulting scatter is shown in Figure \ref{fig:ax_ratio_comparison};} for these untuned models there is an error of $\sim0.1$, consistent with our expected errors (derived in Section \ref{sec:error_estimation}).

We observe a clustering of outlying values around $b/a=0.5$. This is almost certainly due to the drawing tool ellipse having a default axis ratio of 0.5. Where this default is a ``good enough'' fit we hypothesise that volunteers are less likely to modify it, while if it needs to move a long way they find a more refined value. Overall we see that 36\% of all disc components drawn by volunteers were left at the default axis ratio. We recommend that future projects should carefully consider their interface design to minimize this bias (e.g. forcing volunteers to draw both the major and minor axis), however, the fitting process we implement on the aggregate models successfully removes the bias, and the overall scatter does not change significantly.

As we account for light in spiral arms and bars, we expect that disc axis ratios fit by \textit{Galaxy Builder} should be more physical than those from models that do not account for how these non-axisymmetries can bias measurements of ellipticity.

\subsubsection{Comparison to Disc-Bulge models}

A strong motivation for performing multi-component modelling is the desire to measure the fraction of a galaxy's light being emitted by its central components (such as bulge fraction, defined as the ratio of bulge luminosity to total luminosity). \citet{Gao2017:1709.00746v1} demonstrate that modelling secondary central components is essential for recovering an accurate measure of bulge fraction. The difficulty of measuring bulge fraction is further compounded by the complex degeneracies present in even two-component fits, meaning that many gradient-descent based solvers often fail to find the globally optimum solution \citep{profit-paper}, especially when bulge S\'ersic index is a free parameter.

One of the largest catalogues of 2D multi-component fits is \citet{2011ApJS..196...11S}, which performed simultaneous, two-bandpass decompositions of 1,123,718 galaxies in the Legacy area of the SDSS DR7 using \textsc{Gim2D}. Three variations of models were fitted: a pure S\'ersic model, an Exponential disc and de-Vaucouleurs bulge model (hereafter exp+deV), and an Exponential disc and a S\'ersic bulge model (exp+S). Fitting was performed using the Metropolis algorithm, which is resilient to local minima and therefore suitable for the complex likelihood space of galaxy photometric modelling. \citet{2012MNRAS.421.2277L} similarly fitted two models to SDSS main-sample galaxies: an exponential disc and exponential bulge (exp+exp), and an exponential disc and de Vaucouleurs bulge. They used a Levenberg-Marquardt gradient descent algorithm, with initial parameters taken from previous SDSS analysis.

\begin{figure*}
  \plotone{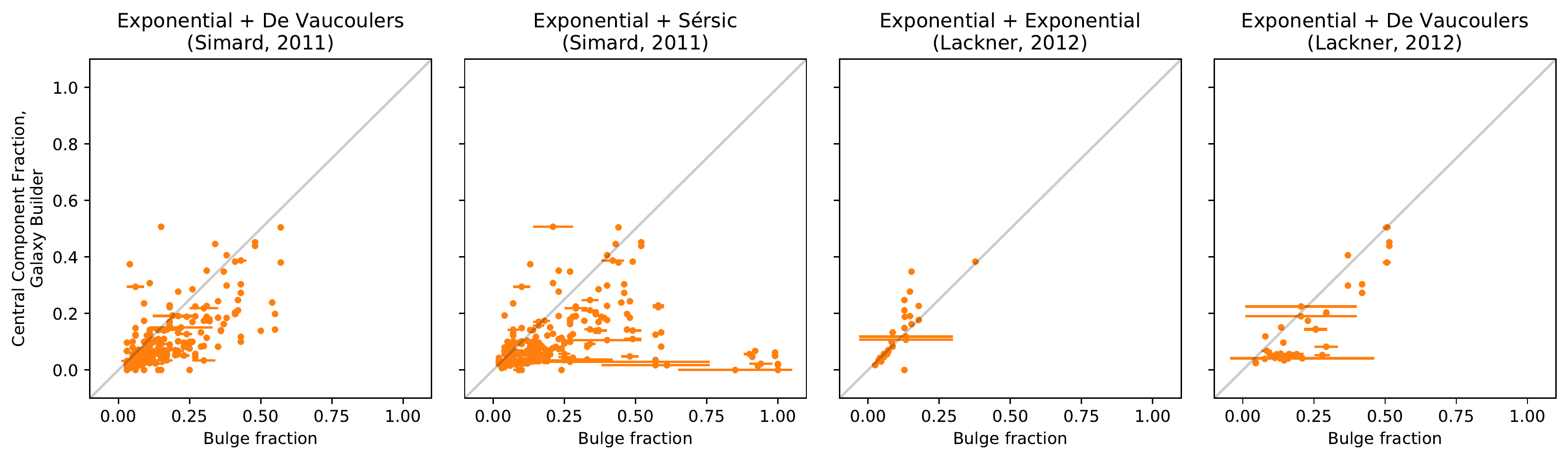}
  \caption{Scatter plots comparing the ratio of flux from central components (bulge and bar) to the total flux between fitted models from \textit{Galaxy Builder} and two-component models in the literature. Our models are broadly consistent with their results, but should be more accurate for complex galaxies, as we account for galaxy bars.}
  \label{fig:bulge_fractions}
\end{figure*}

We compare our central component fraction (the flux of the bulge and bar relative to the total model flux) to bulge fraction from \citet{2011ApJS..196...11S} where their analysis indicated genuine bulge+disc systems ($P_{pS} \le 0.32$). We compare to \citet{2012MNRAS.421.2277L} bulge fractions only when their model selection criteria determined that model was the best-fit model. We see a strong correlation with significant scatter (Figure \ref{fig:bulge_fractions}). The relationship to exp+deV models appears to be less than 1:1, while the relationship to exp+exp models is greater than 1:1, highlighting the dependence of bulge fraction on S\'ersic index. Taking \textit{Galaxy Builder} results as ground truth implies that exp+deV puts too much light into the bulge, while exp+exp puts too little.

The amount of scatter (and lack of consistent 1:1 relationships) between bulge fractions between any two of the published two-component models is comparable to the scatter we see between any one of them and our more complex model. Bulge fractions for complex multi-component galaxies fit with any method should be used with caution.

\begin{figure*}
  \plotone{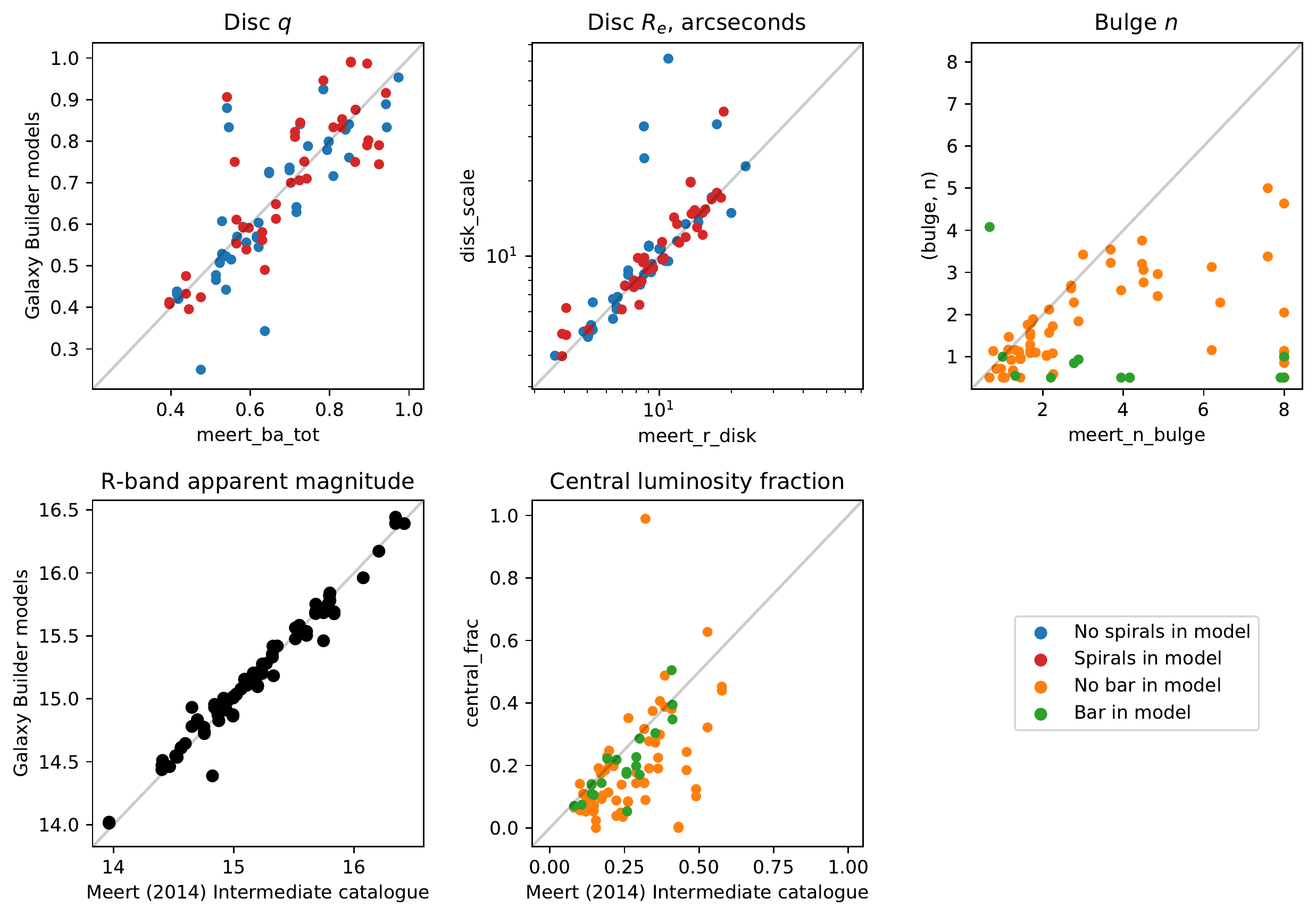}
  \caption{\added{Scatter plots comparing measured model parameters between \citeauthor{2015MNRAS.446.3943M} (\citeyear{2015MNRAS.446.3943M}, x-axis) and \textit{Galaxy Builder} (y-axis). We note that adding spirals to a model does not strongly impact disc parameters, but the presence of a bar has a significant impact on bulge S\'ersic index measurement.}}
  \label{fig:meert-comparison}
\end{figure*}

\added{Another comprehensive catalogue of 2D two-component fits is that of \citet{2015MNRAS.446.3943M}, who fit identical models to \citet{2011ApJS..196...11S} on $\sim 7 \times 10^5$ galaxies imaged by SDSS, using \textsc{Galfit} and \textsc{PyMorph} \citep{2010MNRAS.409.1379V}. They made use of a set of logical filters to distinguish between model fits, allowing them to identify cases where the model did not converge to a physically meaningful result. There is an overlap of 86 galaxy builder galaxy models with their ``intermediate catalogue'', and we see some scatter between measured parameters (see Figure \ref{fig:meert-comparison}). The modelling of spiral arms does not appear to impact measured disk parameters, with disk size and ellipticity showing strong agreement between the catalogues. We see significant scatter in bulge S\'ersic index, especially when a bar is present. Total luminosity is not strongly affected by the addition of detail to the model.}

\subsubsection{Comparison to Disc-Bulge-Bar models}

\begin{figure}
  \plotone{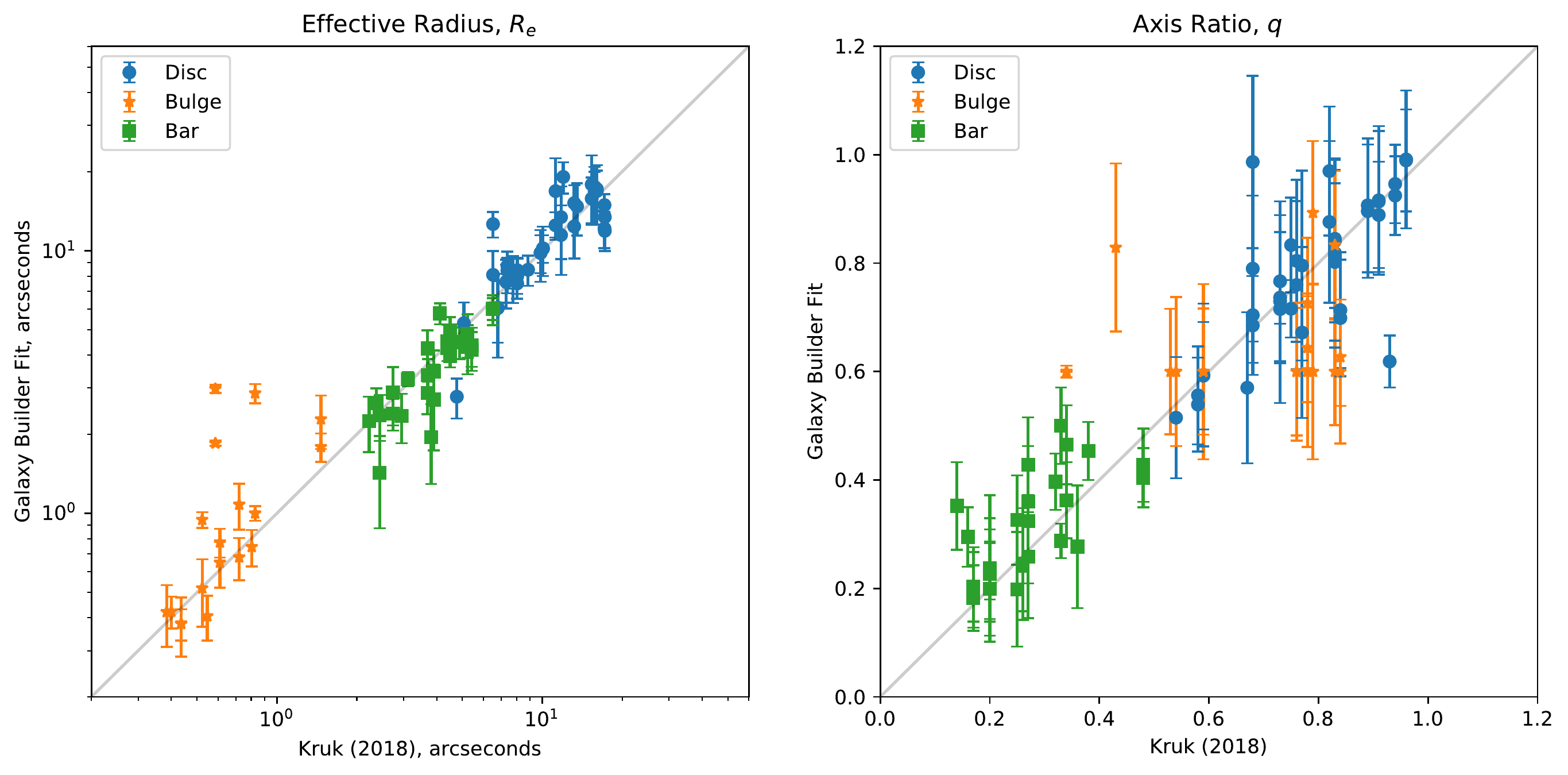}
  \caption{Comparison between \textit{Galaxy Builder} fitted models and the result of 3-component, multi\-wavelength fits performed by \citet{2018MNRAS.473.4731K}. Discs, Bulges and Bars are shown as blue circles, orange stars and green squares respectively. The left panel compares components' effective radii, the right panel compares the component axis ratio. The components match well, with bulges showing the most scatter. Bulges in \textit{Galaxy Builder} fit models often get stuck at the lower allowed value, despite the physically motivated initial conditions.}
  \label{fig:sd_comp_comparison}
\end{figure}

\citet{2018MNRAS.473.4731K} performed multi-component (up to three), multi-band decompositions of a selection of SDSS galaxies, 23 of which were also classified in \textit{Galaxy Builder} (with 16 in the repeated validation subset). Figure \ref{fig:sd_comp_comparison} compares the axis ratios and effective radii of bulges, discs and bars in \citet{2018MNRAS.473.4731K} to those present in the fitted models. We see good consistency in effective radii of all components in the majority of galaxies. There is more scatter in the fit axis ratios of components. In particular, we observe many of the \textit{Galaxy Builder} bulges reaching the imposed lower boundary. Comparing the central component fraction between \textit{Galaxy Builder} models and those in \citet{2018MNRAS.473.4731K}, we see next to no scatter.

\subsubsection{Comparison to Disc-Bulge-Bar-Spiral models}
To the best of our knowledge, no photometric models exist for the \textit{Galaxy Builder} sample that contain spiral arm structure. The closest comparable result is that produced by \citet{Gao2017:1709.00746v1}, however, the galaxies they used are not in the Sloan footprint.

In order to provide a comparison for our novel method of spiral parameter (pitch angle and amplitude) extraction, we compare the result of our galaxy length-weighted pitch angles to the relationship obtained by \citet{Hart2016:1607.01019v1} between GZ2 classification and galaxy pitch angle. Their fit was obtained by using the Zooniverse project \textit{Spiral Spotter} to filter good vs bad spiral arm segments identified using an automated spiral arm detection and fitting tool, \textsc{SpArcFiRe} \citep{Davis2014:1402.1910v1}, whereas \textit{Galaxy Builder} asks volunteers to provide their own opinion on spiral arm number, location and tightness. \textit{Galaxy Builder} pitch angles are within the (large) uncertainties on the \citet{Hart2016:1607.01019v1} fit.

Many researchers (\citealt{Davis2014:1402.1910v1}, \citealt{2019arXiv190804246D} to name a few) have noted that many galaxies show large inter-arm variations in pitch angle, suggesting that obtaining a single value of a galaxy's pitch angle is highly dependent on which arms have been identified. We plan to further explore this issue in future work.

\section{Summary and Conclusions}
\label{sec:conclusions}
In this paper, we present a novel method for modelling of galaxy images, \textit{Galaxy Builder}, which was conceived with the goal of solving the ``quality or quantity" dilemma facing galaxy image modelling, which, despite advances in computation, still typically requires significant human interaction to achieve quality fits. In future work, we use this sample to investigate spiral arm formation mechanisms.

\textit{Galaxy Builder} leverages the power of crowdsourcing for the hardest to automate parts of image fitting, namely determining the appropriate number of model components to include, and finding regions of parameter space close to the global optima.

The use of a small sample of synthetic images to calibrate and test our model clustering and fitting code has demonstrated our ability to recover galaxy morphology in the majority of cases. For example, our spiral arm fitting recovered spiral pitch angles to within 9 deg. This set of 9 synthetic images revealed a systematic tendency for volunteers to incorporate more bulges and fewer bars than necessary for photometric models of strongly barred spirals. Future work might implement an improved clustering algorithm and an improved user interface to address the failures of bar model clustering we observed in a small fraction of galaxies.

Some parameters \replaced{do not cluster}{are not recovered} well (bulge and bar S\'ersic $n$, bar boxiness), we hypothesise that this is because a wide range of values fit the light profile well. As a result, we are unable to obtain reliable physical results with our optimization algorithm (gradient descent-based methods are subject to being trapped in local minima, or not converging for parameters with flat likelihoods). A solution to this would be performing a full Bayesian optimization with priors obtained from volunteer input, or using a more robust algorithm (such as Basin-Hopping; \citealt{1998cond.mat..3344W}). This work is beyond the scope of the current study.

We have demonstrated our ability to obtain physically motivated models with comparable reduced chi-squared values (between 1 and 5) to results in the literature. We obtain errors on parameters where possible through the sample standard deviation of component clusters, which is less likely to be an under-estimate than approximations using the local curvature of the Likelihood-space.

We compare these new models to existing results in the literature. We find good agreement where the models or parameters are comparable, and suggest that where differences are found, \textit{Galaxy Builder} should generally provide superior models because of the more realistic modelling of the galaxy morphologies.

\deleted{We were able to obtain aggregate models for 296 images with an average rate of one galaxy per day, and fit photometric models for 294 images. User experience and task simplification will need to be considered if significantly larger numbers of these models are to be obtained.}

\added{Upcoming survey missions such as LSST \citep{2019ApJ...873..111I} and Euclid (\citealt{2011arXiv1110.3193L}, \citealt{2012SPIE.8442E..0ZA}) present a rich source of astrophysical data. However, the approach detailed in this paper will not be sufficient to deal with the volume of galaxies these surveys will image (twenty billion and two billion respectively, though a large proportion of these will not benefit from detailed photometric modelling). Tools such as \textit{Galaxy Builder} may serve an important role in the generation of training catalogues for scalable machine learning techniques, in an analogous manner to that currently employed for visual morphological classification in Galaxy Zoo: Enhanced \citep{2020MNRAS.491.1554W}.}

\added{We were able to obtain aggregate models for 296 images with an average rate of one galaxy per day, and fit photometric models for 294 images.} At the time of writing and to the best of our knowledge, the number of photometric models obtained here is still significantly larger than the largest sample obtained through purely computational photometric fitting of a disc, bulge, bar and spiral arms in galaxies (10 galaxies, \citealt{Gao2017:1709.00746v1}, who also included rings, disc-breaks and further components).

\added{The software used to generate image cutouts; perform clustering and aggregation of volunteer models, and fit photometric models is available under a GNU general public licence on GitHub\footnote{\url{http://github.com/tingard/gzbuilder_analysis}}. We hope that publishing this code with the paper promotes transparency and accountability in astrophysical software development. All models created as part of the \textit{Galaxy Builder} project will be available on the Galaxy Zoo website\footnote{\url{https://data.galaxyzoo.org}}.}

Any citizen science project is only as good as the volunteers who generously donate their time to it. We were incredibly fortunate to be able to make use of the wonderful pool of volunteers built by the Zooniverse, who in some cases contributed hundreds of detailed galaxy classifications to this project. We are optimistic about the potential of projects like \textit{Galaxy Builder} to dramatically increase the ability of researchers to perform complex, labour-intensive modelling of galaxy photometry, leveraging the power of the crowd to perform the complex tasks best suited to humans, and computer algorithms for the final optimization.

\section{Acknowledgements}
\label{sec:acknowledgements}
\added{We would like to thank our referee for their input; suggesting numerous ways to improve the clarity and content of this paper, and providing avenues of discussion that we had not previously covered. We would also like to thank Ross Hart for providing the target catalogue for the \textit{stellar mass-complete sample}.}

This publication made use of SDSS-I/II data. Funding for the SDSS and SDSS-II was provided by the Alfred P. Sloan Foundation, the Participating Institutions, the National Science Foundation, the U.S. Department of Energy, the National Aeronautics and Space Administration, the Japanese Monbukagakusho, the Max Planck Society, and the Higher Education Funding Council for England. The SDSS Web Site is \url{http://www.sdss.org/}.

This publication uses data generated via the Zooniverse.org platform, development of which is funded by generous support, including a Global Impact Award from Google, and by a grant from the Alfred P. Sloan Foundation. We would also like to thank the 2,340 volunteers who have submitted classifications to the \textit{Galaxy Builder} project, especially user EliabethB, whose presence on the \textit{Galaxy Builder} forum on top of a large number of galaxies modelled has been a huge help.

Montage is funded by the National Science Foundation under Grant Number ACI-1440620, and was previously funded by the National Aeronautics and Space Administration's Earth Science Technology Office, Computation Technologies Project, under Cooperative Agreement Number NCC5-626 between NASA and the California Institute of Technology.

This project was partially funded by a Google Faculty Research Award to Karen Masters (\url{https://ai.google/research/outreach/faculty-research-awards/}), and Timothy Lingard acknowledges studentship funding from the Science and Technology Facilities Council (ST/N504245/1).

\appendix

\section{Model Fitting}

\label{sec:appendix-model-fitting}

Assume Normal priors on component parameters determined from clustering ($\mu_x$, $\mu_y$, $q$, $Re$), with the spread given by the spread in the clustered values. We, therefore, have that our final log-likelihood (to be maximised) is the sum of the Gaussian log-likelihood of the residuals given the pixel uncertainty and the Gaussian log-likelihood of the variation in parameters, given their uncertainty.

The model being rendered is the PSF-convolved sum of the separate components and outputs an ($N_x$, $N_y$) image. The disc, bulge and bar are variations on the boxy S\'ersic profile:

\begin{equation}
I_\mathrm{sersic}(\vec{P}) = I_e \exp\left\{-b_n\left[\left(\frac{r\,(\vec{P})}{R_e}\right)^{1/n} - 1\right]\right\}
\end{equation}

where

\begin{equation}
r\,(\vec{P}) = \left|\begin{pmatrix}
\frac{1}{q} & 0 \\
0 & 1
\end{pmatrix}\begin{pmatrix}
\cos\psi & -\sin\psi\\
\sin\psi & \cos\psi
\end{pmatrix}\left(\vec\mu - \vec{P}\right)\right|_{\ c}.
\end{equation}

The disc is resticted to $n=1; c=2$, bulge to $n\in(0.5, 6); c=2$ and bar to $n\in(0.5, 6); c\in(0.5, 6)$.

The S\'ersic components are actually rendered at 5x the image resolution, and downsampled using the mean pixel brightness. This is a widely used method of approximating the true pixel value, which is an integration over the area of sky inside the pixel: for a pixel of size $(\delta_x, \delta_y)$,

\begin{equation}
I_\mathrm{pix}(\vec{P}) = \frac{1}{\delta_x \delta_y}\int_{-\delta_y/2}^{\delta_y/2}\int_{-\delta_x/2}^{\delta_x/2}\mathrm{d}x\mathrm{d}y\ I_\mathrm{sersic}\left(\vec{P} + \begin{pmatrix}
\delta_x \\
\delta_y \\
\end{pmatrix}\right).
\end{equation}

Spiral arms were restricted to be logarithmic with respect to the inclined, rotated disc. They were rendered in a similar manner to the online interface; using the nearest distance from a pixel to a calculated logarithmic spiral.

An inclined, rotated log spiral requires parameters brightness $I_s$, spread $s$, minimum and maximum $\theta$ ($\theta_\mathrm{min}$ and $\theta_\mathrm{max}$), an amplitude $A$, pitch angle $\phi$, position $\vec\mu$, position angle $\psi$ and axis ratio $q$, where $\vec\mu$, $\psi$ and $q$ are inherited from the disc component.

The distance from a pixel to a logarithmic spiral is given by
\begin{equation}
  D_\mathrm{s}(\vec{P}) = \min_{\theta\in[\theta_\mathrm{min}, \theta_\mathrm{max}]}\left|\left|\vec{P} - \vec\mu - Ae^{\theta\tan\phi}\begin{pmatrix}
       \cos\psi & \sin\psi\\
       -\sin\psi & \cos\psi
       \end{pmatrix}
       \begin{pmatrix}
       q\cos\theta \\
       \sin\theta \\
       \end{pmatrix}
       \right|\right|^{\ 2}.
\end{equation}

In practice the spiral distance was approximated using the distance to a poly-line with 200 vertices, as solving the above minimization for each pixel at each fitting step is computationally intractable. We also adjust $A$, $\theta_\mathrm{min}$ and $\theta_\mathrm{max}$ to account for the rotation of the disc component from its starting value, in order to prevent spirals inadvertently moving far from starting locations for face-on discs (which have poorly constrained position angles). These adjustments are

\begin{equation}
\begin{aligned}
  A' &= Ae^{\Delta\psi\tan\phi},\\
  \theta_\mathrm{min}' &= \theta_\mathrm{min} - \Delta\psi,\\
  \theta_\mathrm{max}' &= \theta_\mathrm{max} - \Delta\psi.
\end{aligned}
\end{equation}

The pixel brightness is then calculated as

\begin{equation}
I_\mathrm{spiral}(\vec{P}) = I_{e,\,\mathrm{disc}}(\vec{P}) \times I_s\exp\left(\frac{-D_\mathrm{s}(\vec{P})}{2s^2}\right).
\end{equation}

For the fit, we parametrize disc $I_e$ as the S\'ersic total luminosity, given by

\begin{equation}
L_\mathrm{tot} = I_e R_e^2\ 2\pi n\frac{e^{b_n}}{(b_n)^{2n}}\Gamma(2n).
\end{equation}

Bulge (bar) $I_e$ is reparametrized as ``bulge (bar) fraction'', which we define as

\begin{equation}
F_\mathrm{bulge} = \frac{L_\mathrm{bulge}}{L_\mathrm{disc} + L_\mathrm{bulge}},
\end{equation}

and is limited to be between 0 and 1. Disc luminosity is allowed to take any value greater than or equal to zero.

Similarly, bulge and bar effective radius are reparametrized as their scale relative to the disc ($R_e = R_e / R_{e,\,\mathrm{disc}}$). Bulge and bar are also restricted to have the same position.


\bibliographystyle{aasjournal}
\bibliography{bibliography}


\listofchanges

\end{document}